\documentclass[11pt]{emulateapj}
\usepackage{apjfonts}

\newcommand{\kms}{\,km\thinspace s$^{-1}$}

\slugcomment{Accepted for publication in the Astronomical Journal}

\shorttitle{SBF distances for Virgo dEs}
\shortauthors{Jerjen et al.}
\begin{document}


\title{Distances, Metallicities, and Ages of Dwarf Elliptical Galaxies in the Virgo Cluster 
from Surface Brightness Fluctuations}


\author{H.~Jerjen}
\affil{Research School of Astronomy and Astrophysics, 
The Australian National University, Mt Stromlo Observatory, Cotter Road, 
Weston ACT 2611, Australia}
\email{jerjen@mso.anu.edu.au}

\author{B.~Binggeli and F.D.~Barazza}
\affil{Astronomical Institute of the University of Basel, 
Venusstrasse 7, CH-4102 Binningen,
Switzerland}
\email{binggeli@astro.unibas.ch and barazza@astro.unibas.ch}

\begin{abstract}
We have employed FORS1 and 2 at the Very Large Telescope at ESO to acquire deep $B$ and 
$R$-band CCD images of 16 dwarf elliptical galaxies in the direction of the Virgo cluster. 
For each dwarf we measure the apparent $R$-band surface brightness fluctuation (SBF) 
magnitude $\overline{m}_R$ and the $(B-R)_0$ colour in a number of fields at different 
galactocentric distances. From the field-to-field variation of the two quantities we 
determine the SBF distance by means of the $(B-R)_0-\overline{M}_R$ relation. 
The derived distances of the dwarfs are ranging from 14.9\,Mpc to 21.3\,Mpc,
with a mean 1$\sigma$ uncertainty of 1.4 Mpc or 8\% of the distance,
confirming that there is considerable depth in the distance
distribution of early-type cluster members.  
For VCC1104 (IC3388) our SBF distance modulus of 
$(m-M)_{\rm SBF}=31.15\pm0.19$ ($17.0\pm 1.5$\,Mpc) is in good agreement 
with the Harris et al.~(1998) result of $(m-M)_{\rm TRGB}=30.98\pm0.19$\,mag ($15.7\pm 1.5$\,Mpc) 
based on HST observations and the tip magnitude of the red giant branch.
Combining our results with existing distances for giant Virgo ellipticals we identify 
two major galaxy concentrations in the distance distribution: a broad primary clump around 
$(M-m)=31.0$\,mag ($15.8$\,Mpc) and a narrow secondary clump around
$31.33$\,mag ($18.5$\,Mpc). An adaptive kernel analysis finds the two 
concentrations to be significant at the 99\% (2.5$\sigma$) and 89\% ($1.6\sigma$) levels. 
While the near-side clump of Virgo early-type galaxies can be associated to
the subcluster centered on M87, the second clump is believed to be mainly due
to the backside infalling group of 
galaxies around M86. \\
The ages and metallicities of the dE stellar populations are estimated 
by combining the observed $(B-R)_0$ colours with Worthey's stellar population synthesis models. 
It appears that the Virgo dEs cover a wider range in metallicity, from [Fe/H]$\approx -1.4$ 
(VCC0815) to $-0.5$ (NGC4415), than Fornax cluster dEs. The derived metallicities place the 
Virgo dEs on the extension of the metallicity--luminosity relation defined by the 
low-luminousity Local Group dEs. The data further suggest an age range from genuinly old 
($\sim 17$\,Gyrs) stellar systems like IC3019 and IC0783 to intermediate-age ($8-12$\,Gyrs) 
dwarfs like NGC4431 and IC3468.
\end{abstract}


\section{Introduction}
The nearby Virgo Cluster is a largely extended and complex structure of over 
1300 galaxies and represents the major feature of the Local Supercluster.  
Based on the pioneering work by de Vaucouleurs (1961), the photographic Las 
Campanas survey by Binggeli et al.~(1985, hereafter the Virgo Cluster Catalog 
or VCC) and a complementary collection of galaxy redshifts (Binggeli 
et al.~1993), a number of subclumps and gravitationally unbound clouds were 
identified in Virgo from imaging and recessional velocity data. Centrally located 
are two separate subclumps each dominated by a giant elliptical galaxy, i.e. M87 and 
M86 (see Binggeli 1999). On a large scale there is another structure along the north-south
axis of the Virgo cluster, defined by the northern M87/M86 subclumps 
(called ``cluster A'' in Binggeli et al.~1987) and the southern 
galaxy concentration around M49, called here the ``M49 subclump'' (= ``cluster B''). 
These subclusters are bound to the south and west by three galaxy clouds named 
W, W' (de Vaucouleurs 1961), and M (Ftaclas et al.~1984).

\begin{figure}
\epsscale{1.0}
\plotone{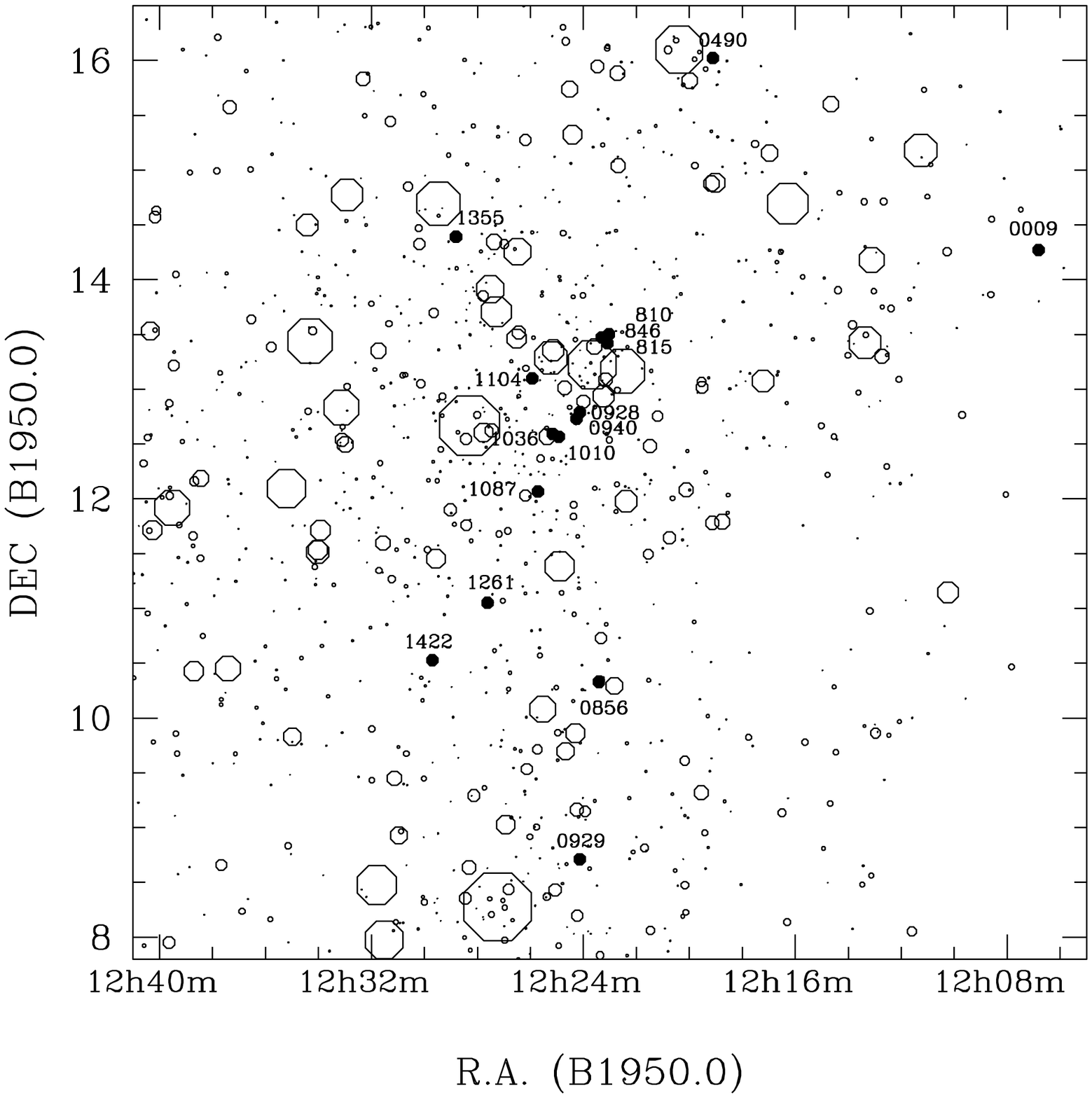}
\caption[]{
A map of all known galaxies in the central region of the Virgo 
cluster, produced from the Virgo Cluster Catalog. The size 
of each symbol is proportional to the galaxy luminosity. Filled circles 
and VCC numbers indicate the locations and names of the  dwarf galaxies 
in our sample. The two most prominent structures known as the M86/M87 
subclumps (cluster A) and the M49 cluster (cluster B) are centered on 
M86 ($12^h 23^m 39.7^s$, $13^\circ 13' 23''$),
M87 ($12^h 28^m 17.6^s$, $12^\circ 40' 02''$) and 
M49 ($12^h 27^m 14.2^s$, $08^\circ 16' 36''$), respectively. }
\label{fig1}
\end{figure}

\placefigure{fig1}

The knowledge of a precise distance to the Virgo cluster or more generally 
a good understanding of its three dimensional structure plays an important role in 
many research areas of extragalactic astronomy. However, despite the large effort to 
resolve these issues over the past decades, the spatial extension of the Virgo cluster 
remained highly uncertain. The reasons are twofold. Firstly, only a relative small number of 
cluster member galaxies had accurate distance measurements. Secondly, the typical 
target objects were spiral galaxies which reside, according to the morphology--density 
relation (Dressler 1980), in the outskirts of galaxy clusters. So it should come as 
no surprise that using accurate distance indicators
like Cepheids and Supernovae of type Ia revealed a significant distance spread among the 
Virgo spirals ranging from 15\,Mpc (eg.~Graham et al.~1999; Saha et al.~2001) to 
25\,Mpc (eg.~Saha et al.~1997). These heavily disputed results simply reflect the 
line-of-sight depth of the large spiral halo of the Virgo cluster and demonstated 
why accurate distance measurements of only a few spirals are of limited use.

Progress towards a precise mapping of the 3D-structure of the Virgo cluster 
region was made only when the numbers of spiral galaxy distances were increased 
(e.g.~Yasuda et al.~1997). Alternatively, fundamental plane distance measurements 
(Gavazzi et al.~1999) and surface brightness fluctuations (Neilsen \& Tsvetanov 2000; 
Tonry et al.~2001) were employed to locate the more centrally concentrated 
early-type giant elliptical galaxies. Similar to the giant  brethren, the less 
luminous and more elusive dwarf elliptical (dE) galaxies are well confined to the 
highest galaxy densities in a cluster (the morphology--density relation for dwarfs, 
Binggeli et al.~1987). But it is their occurrence in large numbers in cluster cores 
(Binggeli et al.~1985; Ferguson \& Binggeli 1994; Jerjen \& Dressler 1997) that makes 
them unique and even more valuable than giant ellipticals. Dwarf ellipticals are the 
only galaxy type that flag the gravitational center(s) in a galaxy cluster 
{\it and} are available in statistically sufficient numbers.

Despite the great potential dEs offer to examine the densest regions of galaxy 
clusters only little work has been done with dEs to date due to the lack of an 
accurate and practical distance indicator. Correlations between global parameters 
of dEs such as the effective surface brightness--luminosity relation (Binggeli 
\& Cameron 1991) or the shape parameter--luminosity relation (Jerjen \& Binggeli 
1997; Binggeli \& Jerjen 1998) have considerable scatter and thus are not reliable 
to measure individual distances. Instead, the resolution 
power of the HST has to be used to resolve the stellar populations of dEs to 
establish distances by means of the tip magnitude of the red giant branch 
(TRGB, e.g.~Karachentsev et al.~2000). While this approach works well for nearby dEs 
it becomes exceedingly difficult (crowding effects) and  expensive 
(long integration times) to obtain good S/N stellar photometry at the 
required limiting magnitude with larger distances. This explains why the 
TRGB method was applied to only one dE (VCC1104 or IC3388) in the Virgo cluster 
(Harris et al.~1998) to date.

Considering the limitation of the TRGB method, another distance indicator has
emerged as a substitute for measuring accurate distances to dEs beyond 10\,Mpc 
from the Local Group. This is the surface brightness fluctuation method based on the discrete 
sampling of an {\it unresolved} stellar population in a galaxy with a CCD detector 
and the resulting Poisson fluctuations in the number of stars within a resolution 
element (Tonry \& Schneider 1988). The method has been extensively tested in
the Sculptor Group (Jerjen et 
al.~1998), Centaurus A group (Jerjen et al.~2000), M81 Group, the Canes Venatici 
cloud and the near field (Jerjen et al.~2001). First results from dwarf ellipticals 
in the more distant Fornax cluster (Jerjen 2003) and Centaurus cluster (Mieske
et al.~2003) were reported recently. The high accuracy of the method ($10-20$\%) opens 
up the possibility to measure precise distances to cluster dEs in a simple and efficient way. 

The goal of this paper is threefold. First, we will measure the Surface Brightness Fluctuation
distances of a sample of 16 dwarf elliptical galaxies in the direction of the Virgo cluster. 
We introduce the dwarf galaxies with their basic properties in \S2. In \S3 we describe the 
observations and data reduction. In \S4 we will carry out the SBF analysis and calculate the 
fluctuation signals. We determine the SBF distances of the sample galaxies in \S5.
Second, we will investigate the distance distribution of the dwarfs in the context of the 
3D-structure of the cluster in \S6. Lastly, we will estimate rough
metallicities and ages for our early-type galaxies in \S7
using the fluctuation magnitudes and B-R broadband colour. Our conclusions are
given in \S8. 

\section{Fundamental Properties of the Sample}
To follow-up the question on the 3D-distribution of the dE population in the Virgo cluster
and to search for the gravitational centre(s) of the cluster, we have studied a  sample 
of 16 early-type dwarfs from the Virgo Cluster Catalog (VCC; Binggeli et al.~1985). As it 
is a main requirement for the successful application of the SBF method, galaxies were 
primarily selected on their morphological appearance, i.e. type ``dE(,N)'' or ``dS0(,N)'' 
and on their apparent size, i.e.~an isophotal radius $r_{B, 25}>30''$ and small ellipticity. 
Within these constraints, dwarfs were chosen in a way to get a good coverage in velocity space 
($-730$\kms$ <V_\odot<$1850\kms) and in the celestial distribution ($12^h 05^m  
< \mbox{R.A.(1950)} < 12^h 40^m$; $+08^\circ 00'< \mbox{Decl.(2000)} 
< +16^\circ 30'$). A few dwarfs like VCC0810 simply came into our sample because 
they were close neighbours of our target galaxy. Finally, we added the nucleated dE 
galaxy VCC1104 (IC3388) to the sample. VCC1104 is the only Virgo dE for which an 
independent distance measurement is available (Harris et al.~1998). 

Fig.~\ref{fig1} shows the galaxy distribution in the central region of the Virgo 
cluster with our sample galaxies highlighted as filled circles. In projection, all 
except two dwarfs are located in cluster\,A (M87/M86 subclump). VCC0929 (NGC4415) 
and VCC0856 (IC3328) are both cluster\,B objects. In Table~\ref{tbl-1} we list the 
fundamental parameters of our dwarfs. The VCC numbers (col.~[1]), other names 
(col.~[2]), and the morphological types (col.~[3]) were taken from Binggeli et al.~(1985.
The J2000 coordinates are give in cols.[4-5] and the cluster region where the galaxy is 
found in col.~[6]. The total $B$ magnitudes (col.~[7]) are taken from Barazza et al.~(2003) 
and the heliocentric velocities (col.~[8]) are from Binggeli et al.~(1985, 1993) and Conselice 
et al.~(2001).

\placetable{tbl-1}

A full account of the photometric properties of the galaxies based on new CCD data, 
including a comprehensive discussion of the characteristics of the surface brightness 
profiles, colour gradients, and structure parameters, is given elsewhere (Barazza 
et al.~2003). We note that the dEs have total absolute $B$ luminosities in the range 
$-17.5<M_{B_T}<-14.5$ and thus are at the bright end of the type-specific luminosity 
function for dEs (Sandage et al.~1985; Jerjen \& Tammann 1997). As such these systems 
are typical cluster dEs with no counterparts known in low density environments i.e.~galaxy 
groups or in the field as satellite galaxies. Only four of our dEs are fainter than 
the M31 companion NGC 205 ($M_{B_T}-15.6$), a dE that is among the brightest 
early-type dwarfs found outside of clusters.
 
\section{Observations}
\subsection{Data acquisition and reduction}
Deep CCD images of the Virgo dEs have been acquired in the $B$ and $R$ passbands in service mode
using the two first units of the Very Large Telescope at ESO Paranal Observatory over a 
period of two semesters: at UT1+FORS1 during an observing run on July 10--14, 
1999 and at UT2+FORS2 during dark time periods in March--May, 2000. 
The detectors of the FORS (FOcal Reducer/Low dispersion Spectrograph) instruments 
are thinned and anti-reflection coated Tektronix (FORS1) and SiTE (FORS2) CCDs 
with $2048 \times 2048$ pixels. By default, service observations were taken in 
standard resolution mode, with a high gain and a pixel scale of $0\farcs2$\,pixel$^{-1}$ 
that yields a field of view of $6\farcm8 \times 6\farcm8$. The CCDs were read out in 
the four-port mode, i.e. four amplifiers read out one quarter of the CCD each. 
Target galaxies were positioned close to the center of one of the four quadrants 
leaving the other three ``empty'' and thus useful as night flats. 
Three exposures of 400--600\,sec durations with slightly different pointings 
were secured in each filter for each galaxy. The Bessell $R$ broadband filter was 
used at UT1+FORS1 while Bessell $B$ and $R_{\rm s}$ filters were employed at 
UT2+FORS2. Table~\ref{tbl-2} summarises the log of the observations.

\placetable{tbl-2}

Template bias frames were subtracted from all CCD images. Twilight sky exposures 
were median-averaged after level normalisation to obtain twilight masterflats. 
Flatfielding with the masterflats did produce no satisfactory results in all cases
and reliable nightflats, which might have worked better, were unfortunately 
impossible to construct from the science frames. To remove the background 
gradients noticeable in some of the images, above all in the $R$ (or $R_s$) frames,
we fitted instead third-order polynomials to the sky background (see Barazza et 
al.~2003 for more details). The systematic uncertainty in the residual images 
due to this sky flattening problem was measured less than 0.1\% over the full 
CCD area in all passbands, with the random errors being much smaller. 

\subsection{Photometric Calibration}
All CCD data were acquired on nights with photometric sky quality. In service 
mode, a series of Landolt (1992) photometric standard stars are observed in various 
passbands under different airmasses every night. Fields which contained saturated 
standard stars when using short exposure times ($\sim 1$\,sec) were reobserved in the 
low gain mode. The analysis of fields taken in both gain modes confirmed the inverse 
gain ratios for the four CCD quadrants given in the FORS1+2 User Manual Issue 2.1 of 
ESO. Finally, we scaled the low gain frames to match the high gain images. In that way,
photometric data over a larger magnitude range became available. From each series of 
standard stars observations we computed photometric zero points, extinction coefficients 
and color terms using the IRAF\footnote{IRAF is distributed by the National 
Optical Astronomy Observatories, which is operated by the Association of Universities 
for Research in Astronomy, Inc., under contract with the National Science Foundation} 
package. 
For the observing period 65 the extinction coefficients and color terms could be 
compared with the values reported by the VLT Quality Control and Trending Services 
(Hanuschik \& Silva 2002\footnote{see also http://www.eso.org/observing/dfo/quality/FORS/ qc/trend\_query\_form.html}) 
finding good agreement:

\begin{eqnarray}
{\rm ESO}&:& k_B=0.25,\quad k_R=0.09,\quad c_{(B-R)}=-0.08 \nonumber\\ 
{\rm Our\,\,study}&:& k_B=0.24,\quad k_R=0.09,\quad c_{(B-R)}=-0.07\nonumber
\end{eqnarray} 

The photometric zero points varied significantly over the observing period 65. 
Therefore, we used the photometric zero points we determined for each 
observing night from our own set of standard stars. Formal photometric 
accuracies in the different passbands are 0.03\,mag (Bessell $B$), 
0.02\,mag (Bessell $R$), and 0.02\,mag ($R_{s}$). QC information for the 
observing period 63 was not available. 

All images of a galaxy were registered by matching the positions of $\approx 100$ 
reference stars on each CCD frame using {\it starfind}, {\it xyxymatch} and 
{\it imalign}. The sky-subtracted images taken in the same passband were 
cleaned from cosmic rays with {\it crreject} and co-added with {\it imcombine}. 
Finally, The resulting master images were flux calibrated. 

\section{SBF analysis}
The $R$-band galaxy images were prepared for the SBF analysis
following the recipes of Jerjen et al.~(2000; 2001). This includes
\begin{itemize}
\item using DAOPHOT~II (Stetson 1987) to clean the images from 
foreground stars, globular clusters, and background galaxies 
brighter than magnitude $m_c=27.0$\,mag ($\approx$500 ADUs in 
$R$ and 600 ADUs in $R_s$).
\item modeling the 2-dimensional galaxy surface brightness 
distribution.
\item subtracting the best-fitting galaxy model from the master image. 
\item noise normalization of the residual image.
\end{itemize}

Between three and seven square subimages (hereafter SBF fields) 
were then defined on a fluctuation image within the 25.5 
$R$\,mag\,arcsec$^{-2}$ isophotal limit thereby avoiding areas 
with previously identified disturbing sources or non-radial 
irregular features such as dust or spiral features  
(Jerjen et al.~2000; Jerjen et al.~2001; Barazza et al.~2002). 
The size of the SBF fields was chosen between $60\times60$ 
and $120\times120$ pixels depending on the apparent size of 
the galaxy. Assuming an average seeing of $0.6$\,arcsec 
(see Table\,\ref{tbl-2}), a field carries the SBF signal from $400$ 
to $1600$ independent points. In total we defined 80 SBF fields 
in our 16 sample galaxies. 

\begin{figure}
\epsscale{0.9}
\plotone{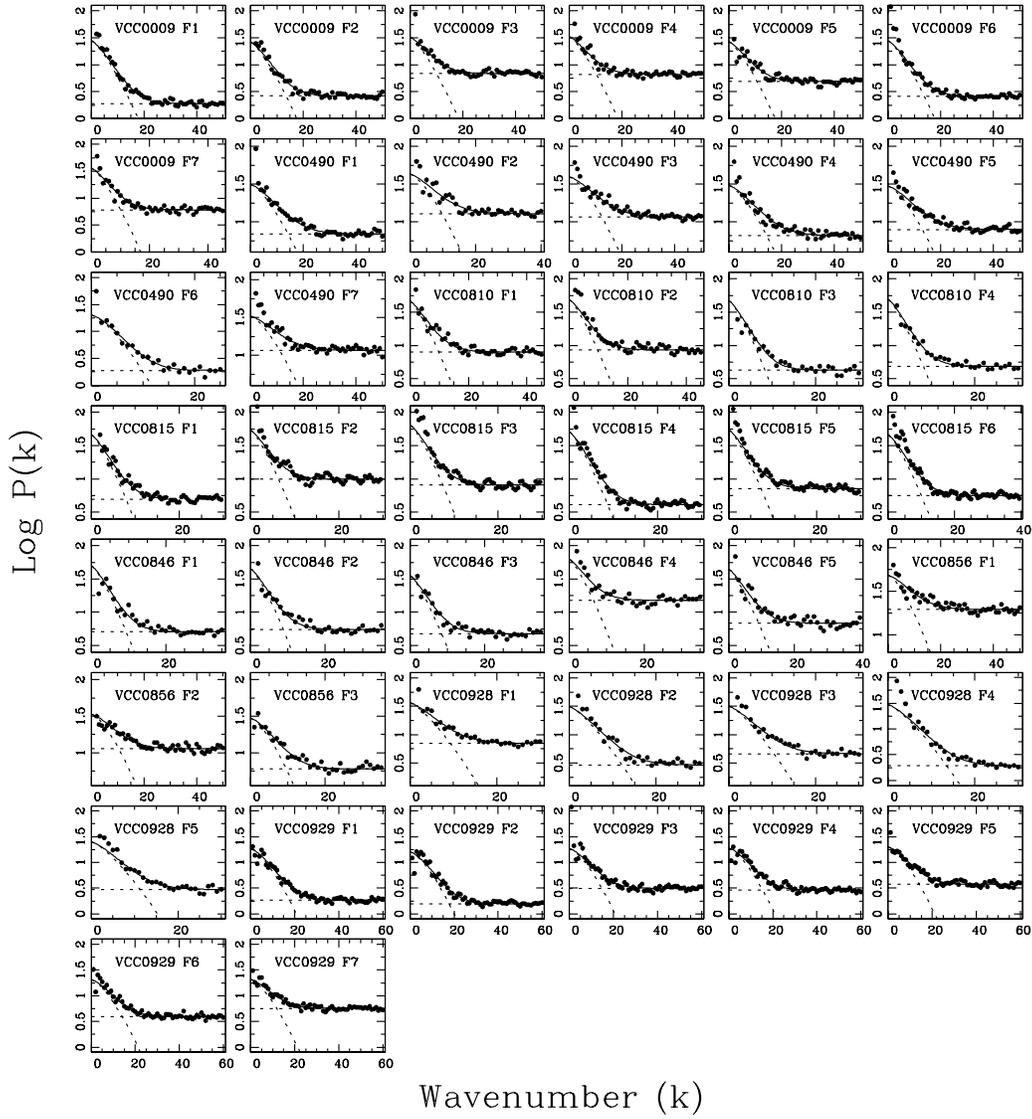}
\caption[]{
Three to seven square fields (subimages) were defined for the SBF analysis
distributed over the surface area of each galaxy. We show here the power 
spectra of the fields analysed in the galaxies VCC0009, VCC0490, VCC0810, 
VCC0815, VCC0846, VCC0856, VCC0982, and VCC929. The SBF field numbers 
are indicated after the galaxy name. Corresponding quantities are listed 
in Table\,\ref{tbl-3}. The observations (filled circles) are well 
fitted by the sum (solid line) of a scaled version of the power spectrum 
of the stellar PSF and a constant (dashed lines).}
\label{psgallery1}
\end{figure}

\begin{figure}
\epsscale{1.0}
\plotone{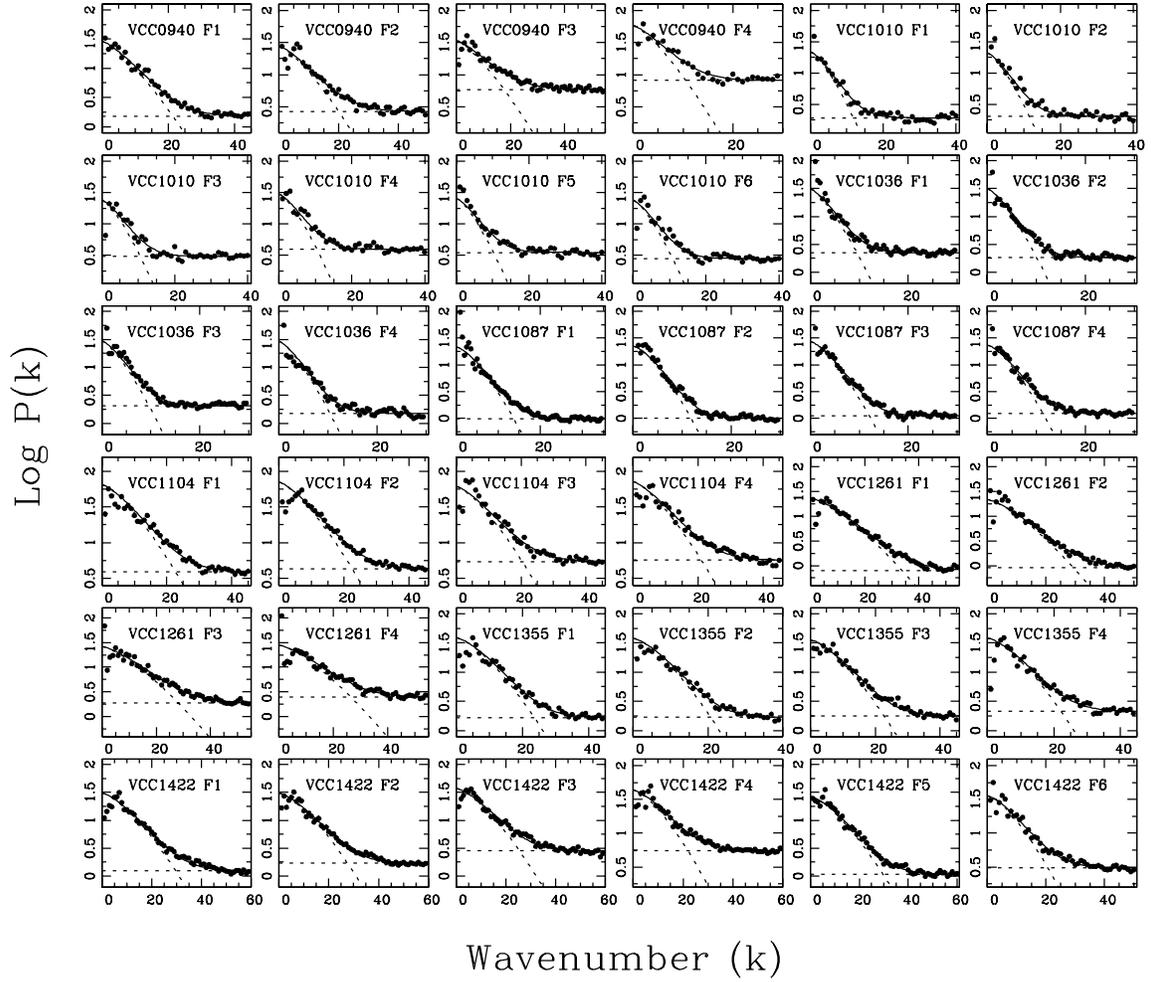}
\caption[]{
The power spectra of all SBF fields analysed in the galaxies VCC0940, VCC1010,
VCC1036, VCC1087, VCC1104, VCC1256, VCC1355 and VCC1422. The SBF field numbers 
are indicated after the galaxy name. Corresponding quantities are listed 
in Table\,\ref{tbl-4}.}
\label{psgallery2}
\end{figure}

\placefigure{psgallery1}
\placefigure{psgallery2}

The SBF fields were Fourier transformed and the azimuthally averaged 
power spectra calculated. From isolated bright stars on the master 
image we determined the point spread function (PSF) profile. We then 
fitted a linear combination of the flux normalized and exposure time weighted 
PSF power spectrum and a constant at the observed galaxy power 
spectrum $\mbox{PS}(k)=P_0\,\mbox{PS}_{\mbox{star}}(k) + P_1$,
demanding a least-squares minimisation. Data points at low spatial 
frequencies ($k\le5$) were omitted as they are likely affected by imperfect 
galaxy model subtraction. Figs.~\ref{psgallery1} and \ref{psgallery2} 
show the power spectrum of each SBF field with the 
best fitting analytic function indicated as solid line. Tables~\ref{tbl-3} and 
\ref{tbl-4} summarize the quantities measured in the SBF analysis: 

Col.~1 -- galaxy name and SBF field number, 

Col.~2 -- pixel size of the SBF field, 

Col.~3 -- magnitude $m_1$ of a star yielding 1 ADU per second on the CCD, 

Col.~4 -- mean galaxy surface brightness within the SBF field in ADU, 

Col.~5 -- sky brightness in ADU, 

Col.~6 -- exposure time normalized amplitude $P_0$ of the best least 
squares fit at wave number $k=0$ with fitting error in brackets, 

Col.~7 -- the scale-free white noise component $P_1$ in the power spectrum,
indicating the ratio of sky to mean galaxy surface brightness within the 
SBF field.
 
The contribution from distant background galaxies that are fainter 
than the cutoff magnitude $m_c=27$\,$R$\,mag to the fluctuation 
signal $P_0$ was estimated with the modified Jensen et al.~(1998) 
equation as discussed in Jerjen et al.~(2001):

$$P_{\rm BG}={{p^2} \over {0.5\ln 10}}10^{(0.8\cdot m_1-22.99)}$$

\noindent where $p$ is the SBF field pixel size in arcsec. 
The measured signal-to-noise S/N$=(P_0-P_{\rm BG})/(P_1+P_{\rm BG})$ 
(Col.~8) in a SBF field is between 2.3 (F1 for VCC0856) and 13.0 (F3 from
VCC1087). The relative contribution of $P_{\rm BG}$ to the 
signal $P_0$ (Col.~9) is of the order of three percent.
 
A rich system of globular clusters (GCs) in the halo of a target galaxy 
is another source of unwanted surface brightness fluctuations. But as 
we pointed out elsewhere (Jerjen 2003) the expected number of GCs in 
bright cluster dwarf ellipticals is quite low and only a minor issue. 
The GC frequency ($S_N$)--luminosity relation for dE,Ns studied in the 
Fornax and Virgo clusters (Miller et al.~1998) predicts 10--40 GCs for 
galaxies in the luminosity range covered by our dwarfs. All GCs would 
be brighter than our cutoff luminosity and  excised during the image 
cleaning process. Therefore, we applied no further correction to the 
measured SBF power. 

\placetable{tbl-3}
\placetable{tbl-4}

\section{Colour--Fluctuation Magnitude Diagrams and SBF Distances}
In the last section we Fourier-analysed the selected SBF fields and 
measured the fluctuation signals. These signals were converted into 
a stellar fluctuation magnitude $\overline{m}_R$ using the formula 
$\overline{m}_R=m_1-2.5\log(P_0-P_{\rm BG})$. Furthermore, the $(B-R)$ 
colour for each field was measured from the cleaned $B$ and $R$ 
master images. Both quantities were corrected for foreground extinction 
using the IRAS/DIRBE maps of dust IR emission (Schlegel et al.~1998). 
The results are listed in the Tables~\ref{tbl-5} and \ref{tbl-6}. 

The power spectrum fitting error is between 2 and 11\%. Other 
sources of minor errors are the PSF normalization ($\sim$2\%), 
the shape variation of the stellar PSF over the CCD area (1--2\%) 
and the uncertainty in the photometric calibration ($0.03$\,mag in $B$, 
0.02\,mag in $R$ and $R_s$). If we further adopt a 16\% error for the foreground 
extinction (Schlegel et al.~1998), the formal combined error for a 
single $\overline{m}_R^0$ measurement is between 0.05 and 0.14\,mag 
(Col.~3). The error associated with the local colour (Col.~4) 
has been obtained through the usual error propagation 
formula from the uncertainties in the sky level determination, 
the photometry zero points, and Galactic extinction. 

The Distance modulus of a program dwarf was then determined by least-squares 
fitting the set of $[(B-R)_0, \overline{m}_R]$ data to the calibration 
equations

\begin{eqnarray}
\overline{M}_R &=& 6.09\cdot (B-R)_0-8.94 \\
&&\quad \mbox{for}\,\,1.10<(B-R)_0<1.50 \nonumber\\
\overline{M}_R &=& 1.89\cdot [(B-R)_0-0.77]^2-1.39\\
&&\quad \mbox{for}\,\,0.80<(B-R)_0<1.35. \nonumber
\end{eqnarray}

These linear and parabolic equations were described and discussed at 
length in Jerjen et al.~(2001) and Jerjen (2003). They were deduced 
from the theoretical $(B-R)-\overline{M}_R$ relation for mainly old,
metal-poor stellar populations as predicted by Worthey's (1994) 
stellar population synthesis models using the evolutionary isochrones 
from the Padova library (Bertelli et al.~1994) and an empirical 
zero point (Jerjen et al.~2001). Fig.\ref{mbarcalib} illustrates 
the calibration diagram showing model $(B-R)$ and $\overline{M}_R$ values 
as a function of age, metallicity, and stellar mixture of the underlying 
stellar population. The solid lines mark the two best-fitting analytical 
expressions (Eqs.1 and 2).

\placefigure{mbarcalib}

To obtain the SBF distance of a galaxy with integrated colour redder 
than $(B-R)_0 \approx 1.35$ (e.g.~VCC0929, VCC1010, and VCC1355) is straightforward 
as the colour--fluctuation magnitude relation is unambiguously represented 
by eq.~1. However, the task is more difficult for galaxies in the colour regime 
where both calibration equations are valid ($(B-R)_0<1.35$). There, the decision 
which of the equations to use has to be made based on the observed trend between colour 
and fluctuation magnitude for the analysed SBF fields. In some cases there is also
independent distance information available that renders one of the two solutions
more likely than the other. The requirement, or desideratum to cover a colour
range as large as possible has to be borne in mind 
during the SBF field selection, but sometimes it is impossible to achieve due
to the lack 
of a colour gradient in the dwarf (Barazza et al.~2003). For galaxies like VCC0815, 
VCC0846, VCC0856, VCC0928, VCC0940, and VCC1422 we found sufficient colour 
range in the SBF fields. The results are illustrated in Fig.~\ref{cmdiagrams} 
where we show that the colour--fluctuation magnitude diagrams of these dwarfs,
with the exception of VCC0928, are 
best fitted with the linear calibration equation. Judged from the 
colour--fluctuation magnitude relation over the observed colour range,
VCC0928 might fit into either the linear or parabolic branch, but the latter
is strongly preferred by the observed velocity of $-254$ km $s^{-1}$
(cf.~Table 1); the linear branch solution would put that galaxy at
too large a distance to be compatible with a negative velocity (cf.~also \S6).

For VCC1104 the colour range covered by the fields is very
small, leaving the calibration ambiguous. However, there is a strong
constraint put by the TRGB distance 
of $(m-M)_{TRGB}=30.98\pm0.20$\,mag ($15.7\pm 1.5$\,Mpc) from the HST
observations of Harris et al.~(1998).  
As the tip magnitude of the red giant branch has been shown to be an accurate distance 
indicator for old and metal-poor stellar populations (e.g.~Da Costa \& 
Armandroff 1990; Gratton et al.~1997; Salaris \& Cassisi 1998), this 
result is useful to identify the correct SBF distance. The correlation of 
$\overline{m}_R$ with the $(B-R)$ colour over the range of galactocentric radii 
probed by the four subframes in VCC1104 is plotted in Fig.~\ref{cmdiagrams}. 
We found a good agreement with the SBF distance modulus $(m-M)_{0,\rm SBF}=31.15\pm0.19$\,mag
($17.0\pm 1.5$\,Mpc) based on the linear branch while the distance modulus of 
$(m-M)_{0,\rm SBF}=30.66\pm0.11$\,mag from the parabolic branch appears too small 
and less compatible with the TRGB result given the quoted uncertainties.
Therefore, we will adopt a true distance modulus of $(m-M)_{0,\rm SBF}=31.15\pm0.19$\,mag
for VCC1104 in the following.  
   
The most interesting situation is found for the dwarfs VCC0009 and
VCC0490. As shown by their colour--fluctuation magnitude diagram in Fig.~\ref{cmdiagrams}, 
the $(B-R, \overline{m}_R)$ data points of each galaxy form two separate groups and the 
least-squares fitting moves them right onto the two calibration branches. This 
is the first time that galaxies have been found with data on both branches. 
Although unusual, this finding is well explained with an age difference between the 
underlying local stellar populations (see \S7). It provides 
first empirical support for the relative offset of the two theoretical branches. 
We further note that VCC0490 has two SBF fields (F5 and F6) with similar $(B-R)$ 
colours but with fluctuation magnitudes that differ by more than 0.4\,mag.
This apparent discrepancy can only be understood by overlaying the data on the 
calibration diagram. The case of VCC0490 nicely demonstrates the importance 
to examine many SBF fields in a galaxy to estimate its distance correctly.

There are four galaxies left, viz.~VCC0810, VCC1036, VCC1087, and VCC1261,
where the colour range is too small and where no independent distance
information as in the case of VCC1104 is available to break the colour
branch ambiguity. Their data can be technically fitted either by the linear 
or the parabolic equation and thus two distances can be inferred for them. 
We have to leave it to future investigations 
to find further support for one of the two distances because of lack of independent
evidence that could help to resolve the ambiguity. We show in 
Fig.~\ref{cmdiagramstwo} the best fits of the dwarfs' colour--fluctuation magnitude 
diagrams to both calibration equations and quote the two distances in 
Table\,\ref{distmods}.

We list in Table~\ref{distmods} the derived SBF distances for all program galaxies.
The mean 1$\sigma$ uncertainty in $m-M$, neglecting the calibration ambiguity in those
four cases, is 0.17 mag, corresponding to 1.4 Mpc or 8\% of the distance. 
The dominant source of this uncertainty is the error in the $(B-R)$ colour
through its strong amplification by the steep colour-fluctuation magnitude relation.
This can be compensated only by having a large number of fields for the SBF
analysis; the more fields the smaller the resulting error. 

\begin{figure}
\epsscale{1.0}
\plotone{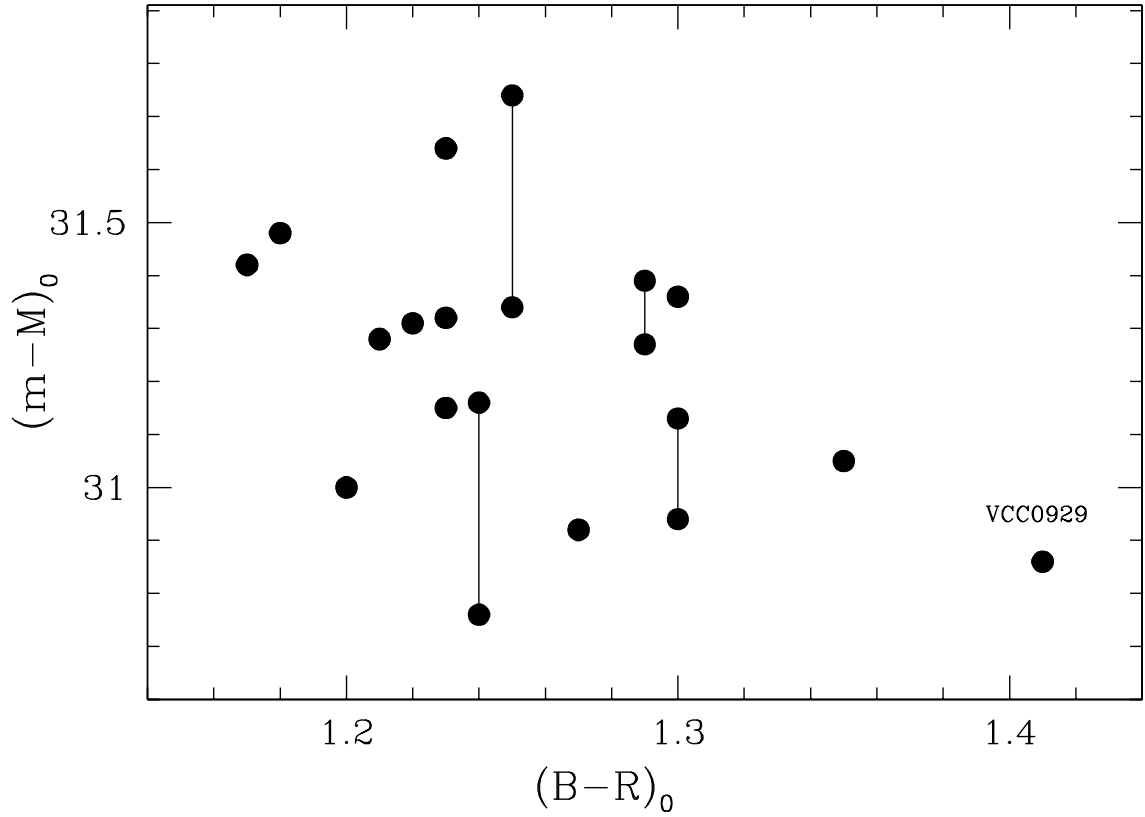} 
\caption[]{
A graph showing the mean $(B-R)_0$ colour versus derived distance moduli
for the 16 sample dwarf galaxies. Points interconnected by a vertical 
line are used when the distance measurement was ambiguous. The impression 
of a slight trend to shorter distances with redder colour is mainly due to 
the short distance measured for VCC0929. However, being a cluster\,B member 
and close to M49, VCC0929 is expected to be slightly more nearby than
the cluster in the mean.   
}
\label{colordistmod} 
\end{figure}

The relatively large galaxy sample with SBF distances available allows us to check 
the quality of our distance calibration. For that purpose we followed Neilsen 
\& Tsvetanov (2000) and tested our data for a possible correlation between the 
colour and distance modulus. Fig.~\ref{colordistmod} shows the distance modulus 
for each galaxy versus the mean color of that galaxy. In the four cases where 
we have two different distance moduli the data points are interconnected with 
a vertical line. At a first glance, there seems to be a slight trend in the data
for redder galaxies to have shorter distances. But looking more closely, this 
impression comes mainly from the relatively short distance of the reddest 
dwarf VCC0929. Do we have reasons to think that VCC0929 is indeed at the near side of 
the Virgo cluster? VCC0929 is a cluster\,B object and in projection close to M49 
(see Fig.\ref{fig1}). As M49 has a reported distance of $31.0\pm0.06$\,mag 
(Neilsen \& Tsvetanov 2000; Tonry et al.~2001) coinciding with the 
mean distance of the M87 subcluster, i.e.~the more nearby of the two major
clumps of cluster\,A (see following section), the short SBF distance of 
$30.86\pm0.14$\,mag for VCC0929 is indeed fully consistent with its location in the 
cluster. Aside from VCC0929, there is certainly no significant colour trend
discernible in Fig.~\ref{colordistmod}.

\placefigure{colordistmod}
\placefigure{cmdiagrams}
\placefigure{cmdiagramstwo}
\placetable{tbl-5}
\placetable{tbl-6}
\placetable{distmods}

\section{Distance and Structure of the Virgo Cluster}
The individual SBF distances of our dEs range from 14.9 to 21.3 Mpc 
(Table~\ref{distmods}, not counting the ambiguous cases). Clearly, 
all of these galaxies are members of the Virgo cluster,
as was expected from their redshift and morphological appearance.
The mean distance modulus, giving each solution of the four ambiguous 
cases half weight, is $\langle (m-M)_0 \rangle$ = 
31.23 $\pm$ 0.06, or $\langle D \rangle$ =
17.6 $\pm$ 1 Mpc. This value is in good accord with the mean Virgo 
cluster distance
derived for giant early-type galaxies from the SBF (Tonry et al.~2001: 
$m-M$ = 31.15, or 17.0 Mpc) and other methods (Neilson \& Tsvetanov 2000,
Kelson et al. 2000), as well as for Virgo spiral galaxies (e.g.~Graham et 
al.~1999; but see Saha et al.~1997 and Ekholm et al.~1999 for dissident
views). However, as the distance distribution of dEs is broad and highly 
non-symmetric (see
below), a discussion of the cluster mean distance is not very meaningful 
anymore;
the derived mean value is obviously highly sensitive to the inclusion or
exclusion of single objects of the sample.

A coarse measure of the distance spread is provided by the standard deviation 
of dE distance moduli:
$\sigma_{\rm obs}$ = 0.245 mag. This has of course to be compared
with the mean distance error for a single dwarf which, as quoted above, is
$\sigma_{\rm dE}$ = 0.169. The true 1$\sigma$ distance dispersion
of the dwarf ellipticals in the cluster is then
$$
\sigma_{\rm cl} = \sqrt{\sigma_{\rm obs}^2 - \sigma_{\rm dE}^2} = 0.177\, 
{\rm mag}\,\,,
$$
or 1.45 Mpc. This corresponds to a total
1$\sigma$ or 2$\sigma$ cluster {\em depth}\/ of the Virgo cluster of 
ca.~3 or 6 Mpc, respectively, in accord with the large distance range 
of 14.9 to 21.3 Mpc
found for the dwarfs (cf.~Table~\ref{distmods}).
How does this compare with the tangential cluster dimension? According 
to Binggeli (1999), the angular width of the Virgo cluster 
is $\approx 8^\circ$ 
(as can also be seen from Fig.~1). Assuming 
spherical symmetry for the cluster and a mean distance of 17\,Mpc, 
we would then expect a ($\approx$ 2$\sigma$) front-to-back 
depth of only 2.4\,Mpc ($0.31$\,mag). Hence this is evidence that the
distribution of Virgo cluster dEs is significantly elongated along the line
of sight. 

There have been previous reports of a {\em very}\/ extended ($8-20$ Mpc), 
cigar-like distribution of Virgo dEs (Young \& Currie 1995). While such a huge
depth in the dE population as claimed by Young \& Currie has been suspected 
to be in large part a 'finger-of-god' effect due to unaccounted-for
observational errors (Binggeli \& Jerjen 1998), it is definitively ruled out
by our data. On the other hand, the slightly prolate distribution of dEs 
found here is in excellent agreement with Neilsen \& Tsvetanov (2000) and 
West \& Blakeslee (2000) 
who used the SBF method to determine the distances to giant ellipticals in
Virgo. These authors 
found a similar cluster depth of $\pm(2-3)$\,Mpc around M87. Moreover, 
Arnaboldi et al.~(2000) 
inferred from the luminosity function of intracluster planetary nebulae that
the front end 
of the Virgo cluster is 14\%--18\% (or $\Delta [m-M]=0.28$ to 0.36)  closer to us 
than the Virgo core region around M87. The filamentary distribution 
of Virgo {\em spirals}\/ has of course been known for a long time. Yasuda et
al.~(1997) have argued that the distribution of spiral galaxies in 
the Virgo cluster is best described as an elongated structure along the 
line-of-sight with a depth of $\pm4$\,Mpc from M87. However, the evidence that
such an elongation, if only a milder one, holds for the early type
galaxies in the core region as well is unexpected and new, and it is nicely 
confirmed by our SBF distances of dwarf ellipticals. 

\placefigure{distmodhisto}

But there is more to the distance distribution than just the dispersion. Consider
Fig.~\ref{distmodhisto}, upper panel, where the distance distribution of our
16 dwarf ellipticals is shown (black histogram, dwarfs with two possible 
distances being weighted 
0.5 at each value). The binned distribution is clearly {\em bimodal}\/ 
with a broad major galaxy concentration at $(m-M)=31.0$\,mag ($15.8$\,Mpc), a 
sharp drop 
at $31.2$\,mag ($17.4$\,Mpc) and a narrow secondary peak at $31.33$\,mag ($18.5$\,Mpc).
These strutural features are further emphasised when we add the SBF results 
for 24 giant ellipticals and NGC4486B from Neilsen \& Tsevtanov (2000) and Tonry et 
al.~(2001) that are found in our surveyed area (dashed histogram). 
The 11 ellipticals 
studied in both references are weighted half at each distance. 

To test the bimodality in the distance distribution against the conservative assumption 
that the entire galaxy sample was drawn from a single Gaussian distribution we used the 
adaptive Kernel method (Vio et al.~1994) employing a Gaussian kernel. Thereby, distances 
for giant elliptical galaxies with two independent measurements (Neilsen \& Tsevtanov 2000; 
Tonry et al.~2001) were averaged while distances for dwarf ellipticals with two possible 
distances were weighted half at each distance. The lower panel in Fig.~\ref{distmodhisto} 
shows the adaptive kernel distribution of the 41 early-type Virgo galaxies as a solid line. 
This line was compared with the result from 1000 simulations of 41 randomly distributed 
objects having the same statistical properties as our sample galaxies, i.e. a 
mean distance modulus of $(m-M)=31.15$ and a standard deviation of $\sigma=$0.4. The mean 
adaptive kernel distribution and the $\pm 1\sigma$ confidence level lines for these 
simulations are shown as dashed and dotted lines. The two galaxy concentrations at 
$(m-M)=31.00$\,mag and $31.33$\,mag were found to be significant at the 99\% (2.5$\sigma$) 
and 89\% ($1.6\sigma$) level, respectively. It should also be noted that any asymmetric
feature in the true distance distribution tends to be washed out by the errors in
the individual distances, i.e.~the two peaks will in reality be much narrrower.

What could this bimodality
mean? It has previously been noted (Binggeli et al.~1993, Binggeli 1999)
that the Virgo cluster core (cluster A) shows a pronounced double structure in
the projected spatial distribution and in the kinematics of the galaxies.
The main, most massive and galaxy-rich clump of cluster A, and of the whole
cluster is clearly centered 
on the supergiant E galaxy M87 (NGC4486). A secondary, less massive and less rich
clump, lying about 1$^\circ$ NW to M87, is centered on the giant galaxy
M86 (NGC4406). Members of this secondary clump are found to have
systematically smaller radial velocities; in particular, all negative
velocities tend to be clustered in the M86 region (Binggeli et al.~1993, Binggeli
1999). This has been interpreted in terms
of an infall, or subcluster merging scenario. Given the observed kinematic 
pattern, the M86 subcluster seems to be falling into the main M87 subcluster   
{\em from the backside}. In consequence, we might expect to see a
bimodality in the distance distribution of Virgo cluster galaxies: 
a somewhat larger, broader population 
peaking at a slightly lower-than-average distance, and a smaller, possibly
narrower population peaking at a slightly higher-than-average distance,
with a separation of 1 to a few Mpc. Not only is this exactly what we see in
Fig.~\ref{distmodhisto}, very intriguingly the distance modulus of the primary,
higher and closer peak, $(m-M)_{\rm peak 1}$ = 31.00, roughly coincides with 
the mean SBF distance 
modulus for M87 measured by Tonry et al.~(2001) and Neilsen \& Tsevtanov
(2000): $(m-M)_{\rm M87} = 31.09\pm0.08$\,mag. Likewise, the position of the 
secondary peak, $(m-M)_{\rm peak 2}$ = 31.33, is perfectly matched by these
authors' measurements for the SBF distance of M86: $(M-m)_{\rm M86} = 
31.31\pm0.19$\,mag. 

The bimodal distance distribution of E and dE galaxies in the core of the 
Virgo cluster is therefore 
naturally explained by the subcluster merging scenario proposed before (Binggeli et
al.1993, Binggeli 1999). The M87 and M86 subclusters seem to be separated by 
$\Delta (m-M) \approx$ 0.3, or $\Delta D \approx$ 2.5 Mpc. With individual
distances and radial velocities available for a fairly large sample of
Es and dEs, we can now also check whether the galaxies show the expected 
infall pattern. For this purpose we have plotted in Fig.~\ref{veldistplot}
distance versus heliocentric velocity for all galaxies condisered.
As the radial velocity dispersion of early-type galaxies in the Virgo cluster
is $\approx 600$\,km\,s$^{-1}$ (Binggeli et al.~1993), we should not expect 
to see any streaming motions in the velocity range $0-2000$\,km\,s$^{-1}$. 
However, there is a clear infall signature traced out
by the galaxies with negative velocities. As mentioned before, most
of these are clustered around M86 (NGC4406), which has a negative velocity
itself. They are on average lying on somewhat larger distances than the 
cluster mean. NGC4419 is an exception, lying in the north of the cluster and 
being classified as Sa
in the VCC. This galaxy likely belongs to a population of surrounding
late-type galaxies that have recently fallen into the cluster (see the
discussion in VCC).
Whether VCC0810, for which two possible distances have been derived
(cf.~Table \ref{distmods}), is approaching the cluster from behind for the 
first time or has already fallen
through it while still maintaining such a large relative (negative) velocity,
is difficult to judge without detailed modeling. The long distance of VCC0810
would in any case appear more likely now, as it is perfectly in accord, 
like the distances and velocities of M86, VCC00928, VCC0815, and VCC0846,
with the virgocentric infall model of Kraan-Korteweg (1986).    

\placefigure{veldistplot}

\section{Estimates of Metallicities and Ages of the dE Stellar Populations}
The metallicities of dE galaxies are basically unknown. Rakos et al. (2003) 
estimated a [Fe/H] range from $-2$ to solar with a mean value around 
$-0.75$. The primary problem in obtaining accurate information about the 
stellar populations in dEs is because their surface brightnesses are very 
low, making spectral observations difficult regardless of distance. In the 
light of those technical obstacles, the much easier to measure field-to-field 
variation of the SBF magnitude and $(B-R)_0$ colours across a galaxy 
may serve as a valuable diagnostic tool to get rough metallicities and ages 
of the unresolved stellar populations. 

Given, however, that the age-metallicity degeneracy steadily increases
for populations younger than 8 Gyrs (i.e. $[(B-R),\overline{M}_R]$ data 
points of models move down along the linear branch in Fig.~\ref{metalmagplot}), 
it is essential to base our estimates on age constraints for early-type galaxies as 
provided by studies using more accurate methods. Stellar absorption-line 
indices were measured by Caldwell et al.~(2003) to disentangle the effects of 
age and metallicity in a sample of Virgo early-type galaxies. The strong 
Balmer lines found in these dwarfs were interpreted within the Lick 
index system as primarily being caused by young age, rather than by low 
metallicity. Derived galaxy ages ranged from 1--16 Gyrs.This interpretation 
stands in contrast to Maraston \& Thomas (2000, 2001), who showed that 
a mix of an old metal-rich and an old metal-poor component can produce 
strong Balmer and metal lines without invoking a young population. 

A full investigation of this issue is beyond the scope of this paper 
but we want to recall instead another result from age-sensitive narrow-band 
photometry in the modified Str\"omgren filtersystem by Rakos et  al.~(2001). These 
authors derived a mean age of $10\pm1$\,Gyrs for a sample of 27 Fornax 
dwarf ellipticals and thus provide observational support for a lower age limit of 
$\approx8$ Gyrs for luminous early-type dwarf galaxies in nearby galaxy clusters.

If the main age of the stellar populations in our dEs is 8 Gyrs or older, as we assume 
in the following,  $R$-band fluctuations and $(B-R)$ colours depend mostly on 
metallicity as both quantities closely track the temperature of the RGB whose 
colour is governed by metallicity. This relation allows a direct comparison 
of the colour range covered by the SBF measurements for an individual 
dwarf with the corresponding metallicity range. 

\begin{figure}
\epsscale{0.5}
\plotone{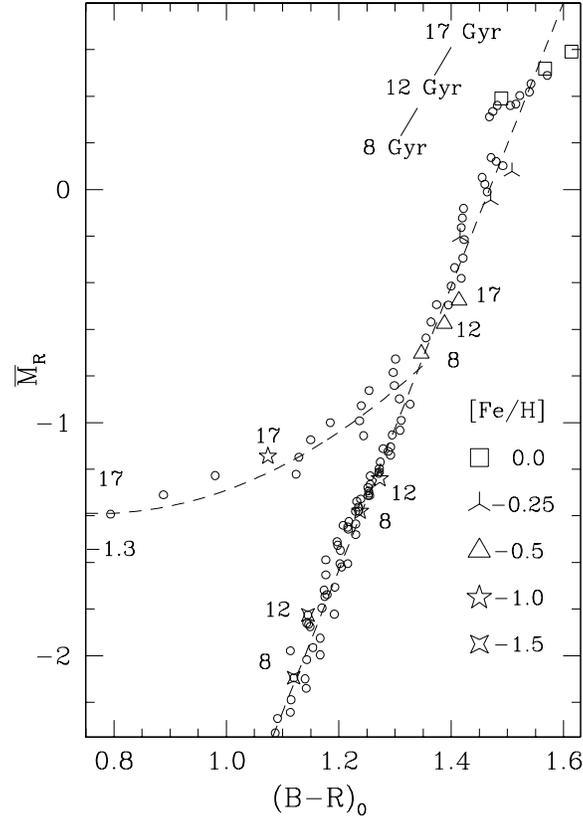}
\caption[]{
SBF calibration diagram showing the Worthey+Padova model predictions of a 
$(B-R)_0$ colour--$R$-band fluctuation luminosity relation for a grid of 
stellar populations (open circles and symbols): a set of single burst populations that 
covers the \{age=8, 12, 17 Gyr\} $\times$ \{[Fe/H]=$-1.7$, 
$-1.6$, ..., $-1.0$, $-0.5$, $-0.25$, $0$\} parameter space 
(with [Fe/H]$\geq-1.3$ in the case of 17\,Gyr due to model limitations)
and a set of composite populations where the previously defined populations 
were mixed at the 10, 20 and 30\% level (in mass) with a second generation 
of 5\,Gyr old stars with solar metallicity. Also shown are two solid lines 
representing the best least squares fits to the two branches exhibit by the 116 
model points.
 The parabolic branch stretching from $0.85< (B-R)< 1.35$ is entirely defined 
 by single-burst, old ($>12$\,Gyr), metal-poor ([Fe/H]$<-0.5$) 
 stellar populations. Slightly younger (8--12 Gyr), more metal-rich single-burst
 populations and mixed populations with a second burst of star formation fall onto 
 the linear branch from $1.0 < (B-R) < 1.5$. A colour independent offset of $0.13$\,mag 
 was applied to the original model data to account for the empirical zero point found 
 in Jerjen et al.~(2001).
}
\label{mbarcalib}
\end{figure}

\begin{figure}
\epsscale{1.0}
\plotone{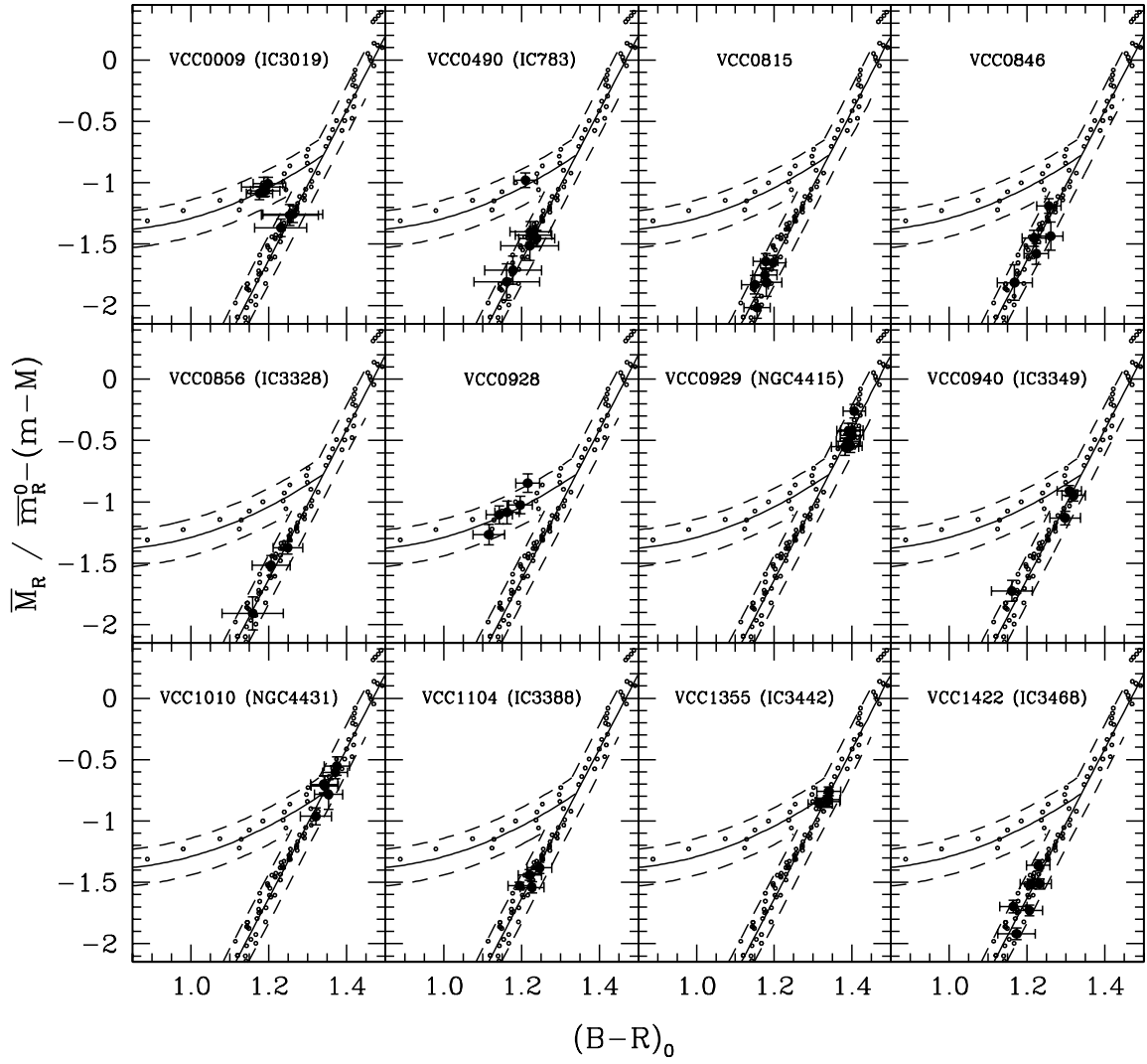}
\caption[]{The colour-fluctuation magnitude relation for all dwarfs with unambiguous
distance. The Fourier analysis of all fields of each
sample galaxy yielded a set of apparent fluctuation magnitudes as a function 
of dereddened local (B-R)$_0$ colour. The data of each galaxy have been 
shifted along the magnitude axis to get the best fit at one of the loci of absolute fluctuation 
magnitudes (small circles and solid lines) computed with Worthey's models and 
using the Padova isochrones with an empirical zero point (see J2003). 
The derived distance moduli are 
listed in Table~\ref{distmods}. The dashed lines above and below 
the branches indicate the $\pm 0.2$\,mag strip that envelopes the intrinsic
scatter from different models and the observed scatter in the fluctuation 
measurements.
}
\label{cmdiagrams} 
\end{figure}

\begin{figure}
\epsscale{1.0}
\plotone{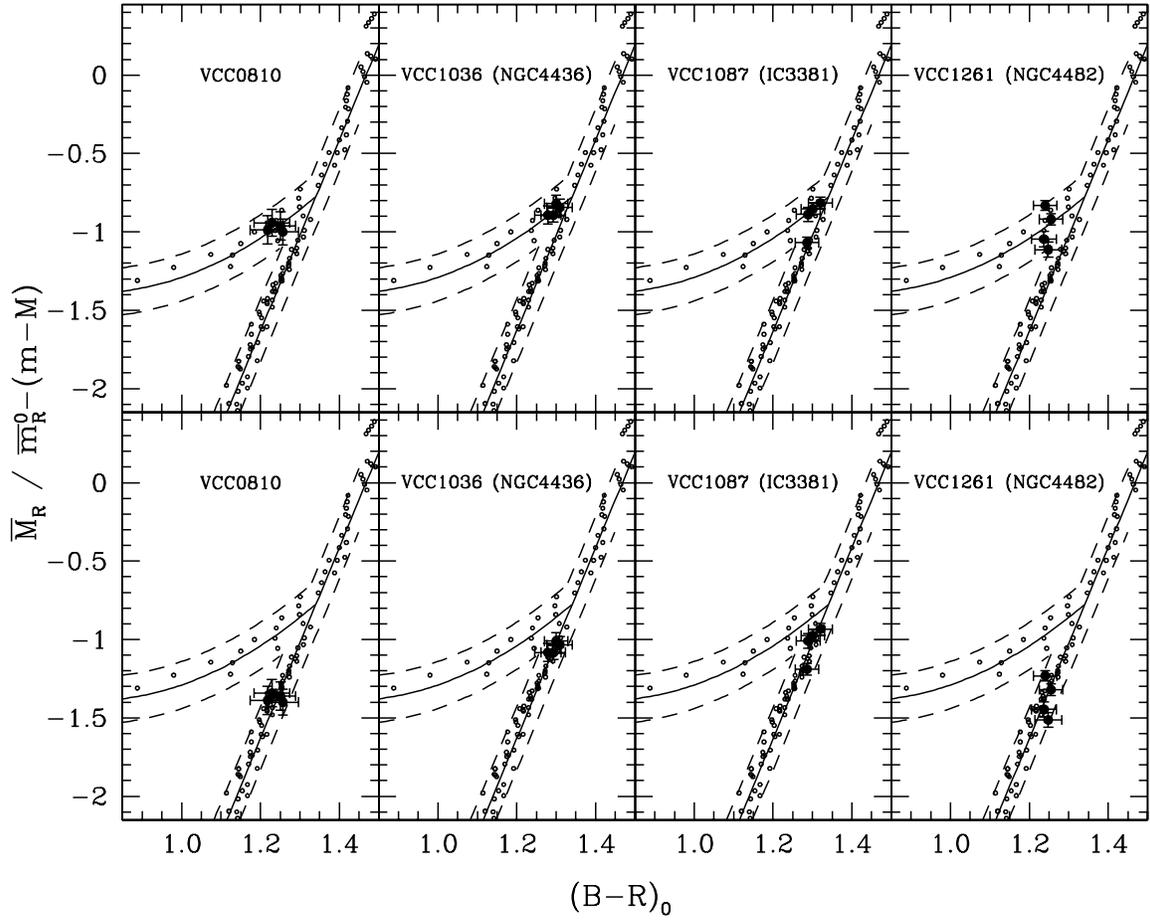}
\caption[]{The same as Fig.\ref{cmdiagrams} but for dwarfs with ambiguous
distances. Due to a small color range, the data can be fitted by both calibration equations. 
The fits producing the shorter distances are shown in the top row while the fits 
associated to the longer distances are presented in the bottom row (see Table \ref{distmods}). 
}
\label{cmdiagramstwo} 
\end{figure}

\begin{figure}
\epsscale{0.6}
\plotone{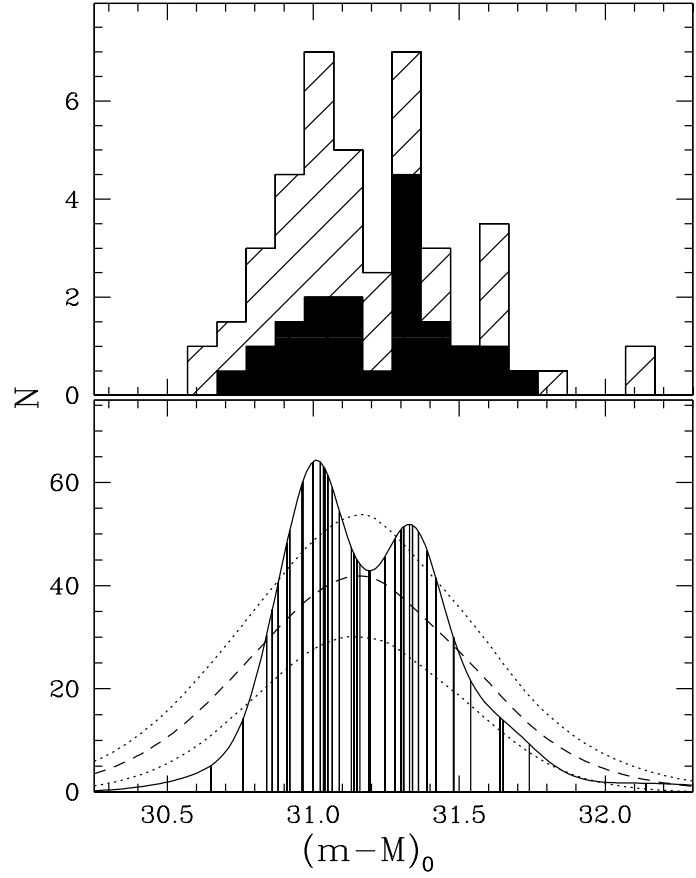}        
\caption[]{
(Top panel) Distribution of the SBF distance moduli of 41 early-type 
dwarf (solid) and giant (dashed) galaxies in the Virgo cluster shown
as a histogram binned in intervals of 0.1\,mag. Data for the giant
ellipticals were taken from Neilsen \& Tsevtanov (2000) and Tonry et 
al.~(2001). The binned distribution shows evidence for a bimodality 
with peaks at $(M-m)_0\approx 31.0$ and $31.3$. (Lower panel) Test of 
bimodality for the galaxy distribution (solid lines) obtained with the 
adaptive Kernel method. Mean values (dashed line) and $\pm 1\sigma$ 
confidence level lines (dotted lines) for 1000 simulations of a Gaussian 
distribution with the same mean (31.15) and standard deviation (0.4) 
as the galaxy data are shown. The two concentrations with exact positions 
at $(M-m)_0=31.0$ and $31.33$ are significant at the 2.5 and $1.6\sigma$ 
level, respectively. 
}
\label{distmodhisto} 
\end{figure}

\begin{figure}
\epsscale{0.6}
\plotone{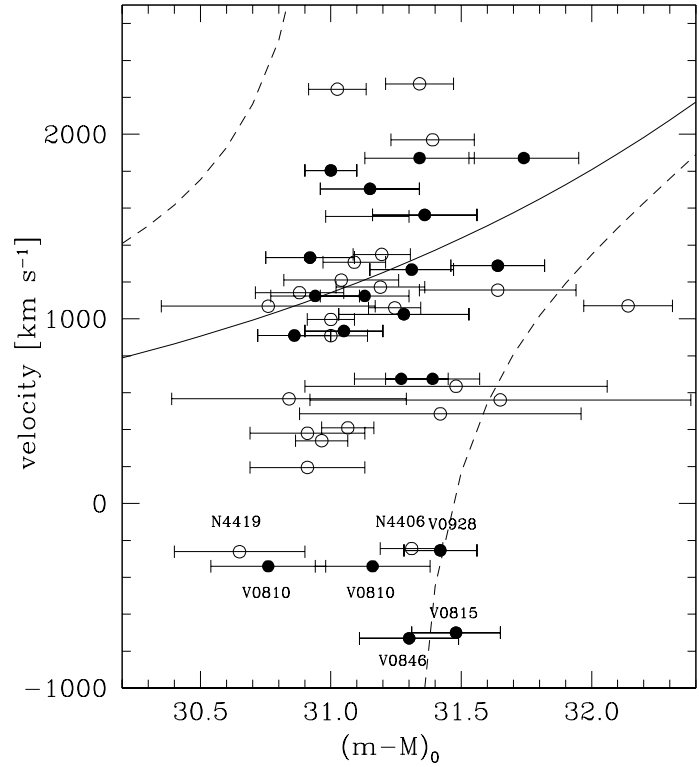}         
\caption[]{
Distance as a function of heliocentric velocity. The filled circles represent 
the dwarfs with SBF distances from this study and the open circle the giant 
ellitpicals with SBF distances from Neilsen \& Tsvetanov (2000) and Tonry et 
al.~(2001). The solid line shows the quiet Hubble flow for 
$H_0=70$\,km\,s$^{-1}$\,Mpc$^{-1}$.
A galaxy falling through the cluster center ($D=17$\,Mpc) along the line-of-sight
would move on one of the dashed lines. These are based on the virgocentric infall model
of Kraan-Korteweg (1986) and a Local Group infall velocity of $v_{\rm LG}=220$\kms. 
The galaxies with negative velocities, including NGC4406 (M86) but with the exception 
of the Sa galaxy NGC4419, follow the expected infall velocity pattern. 
VCC0810 is shown twice due to two possible distances but its long distance 
is more likely to be correct (see text). 
}
\label{veldistplot} 
\end{figure}

\begin{figure}
\epsscale{0.8}
\plotone{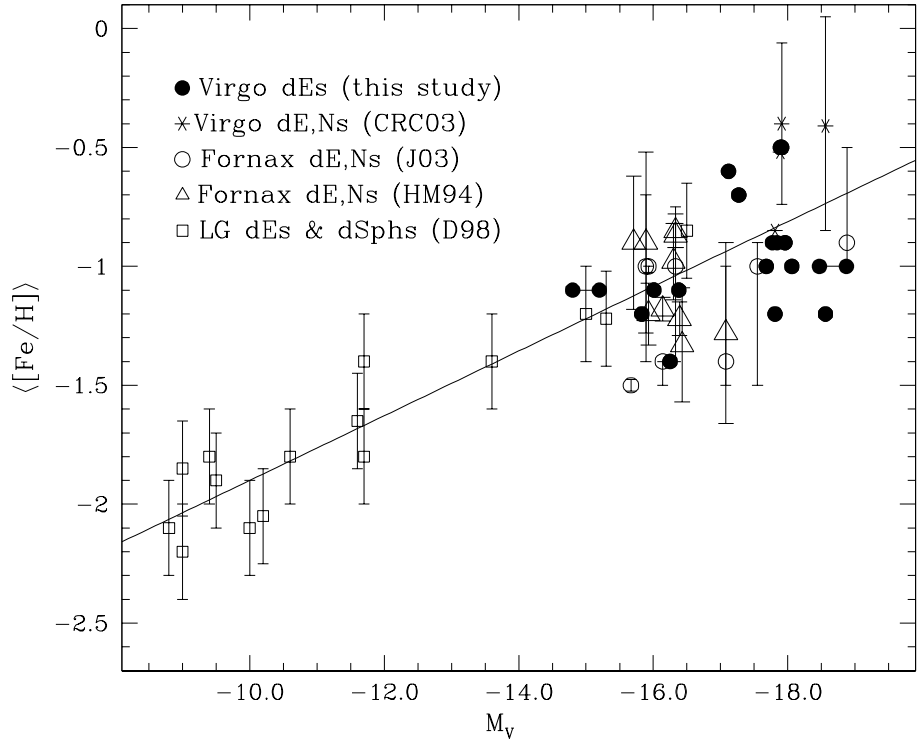}         
\caption[]{
The median metallicities plotted as a function of absolute $V$-band luminosity 
for our sample Virgo dE,Ns (filled circles). The open circle are metallicities 
for eight Fornax dEs derived in a similar way from SBF results (Jerjen 2003). 
The stars, open triangles and squares are the corresponding data for a sample four 
dEs in the Virgo cluster (Caldwell et al.~2003), 10 Fornax dEs (Held \& Mould 1994),
and the dEs and dSphs in the Local Group (da Costa 1998). The error bars are 
taken from the respective reference. The solid line ist the best-fitting
line at the Local Group data and the brighter Fornax dEs 
}
\label{metalmagplot} 
\end{figure}

More quantitatively, we compared each set of observed $[(B-R),\overline{M}_R]$ 
data points with the predictions of a grid of synthetic stellar populations 
(Fig.\ref{mbarcalib}): single-burst populations covering the \{age=8, 12, 17 Gyr\} 
$\times$ \{[Fe/H]=$-1.7$, $-1.6$, ..., $-1.0$, $-0.5$, $-0.25$, $0$\} parameter 
space (with [Fe/H]$\geq-1.3$ in the case of 17\,Gyrs due to model limitations).
Taking into account that local dwarf ellipticals show quite a diversity of evolutionary 
histories with multiple epochs of starformation (e.g. Da Costa 1998; Grebel 1998)
we also added a set of composite models where the previously defined single-burst 
populations were mixed at the 10, 20 and 30\% levels (in mass) with a second, intermediate-age
generation of 5\,Gyrs old stars with solar metallicity. The $(B-R)$ and $\overline{M}_R$ values 
for each model were computed with Worthey's on-line model interpolation 
engine\footnote{http://199.120.161.183:80/$\sim$worthey/dial/dial\_a\_pad.html} 
with a standard Salpeter IMF, using the evolutionary isochrones from the Padova 
library (Bertelli et al.~1994) and the empirical zero point from Jerjen et al.~(2001).

The model values are graphically presented in Fig.~\ref{mbarcalib}. Single-burst, 
old, and metal-poor stellar populations have fainter and bluer $R$-band fluctuations. 
Their data define the parabolic branch while more metal-rich and/or composite 
single-burst populations with a minor contribution of intermediate-age stars make up 
the linear branch. The plot symbols along the linear branch in 
Fig.~\ref{mbarcalib} covering the $8-17$\,Gyrs age range indicate that this part of the 
colour--fluctuation luminosity relation, although degenerated to some level, is primarily 
a metallicity sequence. 

By comparing the location of the SBF data for each dE in Figs. 6 and 7 
with Fig.~\ref{mbarcalib} we get rough estimates of the metallicities. 
Table \ref{metallist} summarises the results and a brief discussion of 
each dE is given here:

\placetable{metallist}

{\small

{\bf VCC0009}: -- appears to be among the purest breed of our sample. 
The locations of the SBF data on both calibration branches suggest  
star formation episods at 17 and 8--12 Gyrs. The  metallicity of 
this major population is [Fe/H]$\approx -1.0$. The data is further
consistent with a pollution at the $\approx 10$\% level with a more metal-rich 
population of intermediate age stars (5 Gyrs old). 

{\bf VCC0490}: -- similar to VCC0009, this dwarf has SBF data points on the parabolic 
branch and thus appears to have a genuine old stellar population (17 Gyrs) with 
a metallicity of [Fe/H]$\approx -1.0$ and a second population (8--12 Gyrs) with 
similar metallicity. Traces of spiral arms found in the $R$-band image of this galaxy 
(Jerjen et al.~2001; Barazza et al.~2002) may be evidence for an even 
younger stellar population. We note that the central part of the dwarf with the spiral
structure was not used in the SBF analysis (see \S 4).

{\bf VCC0810}: --  depending on whether the data is on the parabolic or linear 
branch (see Fig.~\ref{cmdiagramstwo}) we find the following solutions for the 
stellar population. {\it Parabolic branch}: main population has an age of 17 Gyrs 
with $-1.2<$[Fe/H]$<-1.0$ and a 20\% contribution of a 5 Gyrs old population with 
solar metallicity. {\it Linear branch}: 8--12 Gyr old population with metallicity $-1.0$.

{\bf VCC0815}: -- is together with VCC0928 the bluest dE in our sample (B--R=1.15). 
Because the SBF data is located on the linear branch it is also the most metal-poor 
dwarf with a metallicity of [Fe/H]$\approx-1.4$. The main stellar age 
is between 8 and 12 Gyrs. The presence of a second stellar population remains 
undetermined.

{\bf VCC0846}:  -- The metallicity of the stellar population (8--12 Gyrs) is in the range 
of $-1.4<$ [Fe/H] $<-1.0$. The presence of a second stellar population remains 
undetermined.

{\bf VCC0856}:  -- The metallicity of the main stellar population aged 
between 8 and 12 Gyrs is the range of $-1.4<$[Fe/H] $<-1.0$. Caldwell et al.~(2003) derived 
for VCC0856 a much younger age of $3.3-5.2$\,Gyrs and a 0.6\,dex higher metallicity 
($-0.85$ to $-0.43$). The unique spiral structure observed in this early-type 
dwarf (Jerjen et al.~2000) could be optical evidence for recent starformation 
and thus a population of younger stars in this galaxy. We note that the central part 
of the dwarf with the spiral structure was not used in the SBF analysis (see \S 4).

{\bf VCC0928}:  -- The location of the SBF data on the parabolic branch implies a major 
population of 90\% of 17 Gyrs old stars (metallicity range of $-1.3<$[Fe/H]$< -1.0$) with 
10\% of 5 Gyr old, more metal-rich stars.

{\bf VCC0929}:  -- is the most metal-rich dE ([Fe/H]$\approx-0.5$) in our sample
as estimated from $(B-R)_0\sim 1.40$. Caldwell et al.~(2003) derive a similar metallicity 
of [Fe/H]$=-0.52$ and an age of the stellar population of $9.6$\,Gyrs. This 
is in good agreement with the lower age limit (8\,Gyrs) we assumed for 
our metallicity estimates. 

{\bf VCC0940}:  -- The metallicity of the main stellar population is [Fe/H]$\sim-0.7$ 
with an assumed stellar age of 8--12 Gyrs. 

{\bf VCC1010}:  --  Metallicity of the 8--12 Gyrs old stellar population is 
[Fe/H]$\sim-0.5$. A younger population may be present as indicated by the spiral 
arms (Barazza et al.~2002). We note that the area with the spiral structure was not used 
in the SBF analysis (see \S 4). Caldwell et al.~(2003) derived 
a slightly younger age of $5.3-8$\,Gyrs but with a similar metallicity $-0.40\pm0.34$. 

{\bf VCC1036}:  -- Depending on whether the data is on the parabolic or linear 
branch we find the following solutions for the stellar population. 
{\it Parabolic branch}: age 17 Gyrs with $[Fe/H]\approx 1.0$, or {\it linear branch}: 
[Fe/H]$\sim-0.8$ and a typical stellar age of 8--12 Gyrs. The independent 
measurements by Caldwell et al.~(2003) give an age of $4.6\pm6.3$\,Gyrs and 
a metallicity $-0.34\pm0.39$, favouring our latter results and thus 
supporting a distance modulus of $31.13\pm0.17$ for this dwarf.

{\bf VCC1087}:  {\it Parabolic branch}: age 17 Gyrs with 
[Fe/H]$\sim1.0$, or {\it linear branch} [Fe/H]$\sim-0.8$ and stellar 
age of 8--12 Gyrs.

{\bf VCC1104}: -- The SBF data suggests a metallicity of $-1.1$, assuming an age 
of 8--12 Gyrs.   

{\bf VCC1261}: -- Similar to VCC0810 we estimate, depending on which calibration 
branch is used, {\it parabolic branch}: age 17 Gyrs with $-1.2<$[Fe/H]$<1.1$ 
and 20\% of a 5 Gyrs old population with solar metallicity, or {\it linear 
branch}: a slightly more metal-rich ([Fe/H]$\approx -1.0$), 8--12 Gyrs old population.

{\bf VCC1355}:  -- SBF data is located at the intersection of the two branches. 
Independent of the stellar age, but assuming it to be older than 8 Gyrs, we derive a 
metallicity of $-0.6$. 

{\bf VCC1422}:  -- Metallicity of the stellar population is the range of $-1.4<$[Fe/H]
$<-1.0$ with an assumed stellar age between 8--12 Gyrs. Caldwell et al.~(2003) derive 
a metallicity of [Fe/H]$=-0.41\pm0.44$ and an age of the stellar population of 
$4.2\pm4.5$\,Gyrs. The presence of a younger population may be indicated 
by the observed small central edge-on disk (Barazza et al.~2002). But we note that 
this part of the dwarf was not used in the SBF analysis (see \S 4).
}

The broadband colour based metallicities for our Virgo dEs range from 
[Fe/H]$=-1.4$ to $-0.5$. The mean of 0.95 is 0.2 \,dex more metal-rich than the observed mean for a 
sample of eight luminous dEs in the Fornax cluster (Jerjen 2003). The latter sample 
contains none of the metal-rich dEs as VCC0929 and VCC1010 and is centred 
at a slightly fainter absolute magnitude. We calculated the absolute $V$ magnitudes 
for the Virgo dEs (Table \ref{metallist}) from their $B$ magnitudes (Barazza et al.~2003), 
adopting a mean $B-V=0.72$ colour (Bothun et al.~1989), and employing the newly 
derived distance moduli. Fig.~\ref{metalmagplot} shows the location of the Virgo dEs 
in the metallicity-$M_{V}$ plane. The Virgo dwarfs share a common region with 
the mentioned Fornax dEs in the interval $-15.7<M_{V}<-18.9$. The points scatter 
nicely around the metallicity-luminosity relation as set out by the low-luminous dEs and 
dSphs in the Local Group (da Costa 1998) and 10 Fornax dEs (Held \& Mould 1994) 
supporting qualitiatively our results. Also included in the figure, 
for comparison, are the metallicites from Caldwell et al.~(2003) for the four common 
dwarfs VCC0856, VCC0929, VCC1010, and VCC1422. It is not surprising 
that their data preferentially occupy the region above the metallicity-luminosity relation
since higher order Balmer lines were used as age indicators. This 
leads to younger ages ($< 8$\,Gyrs) and higher metallicites as pointed 
out by Rakos et al.~(2001).

Clearly, our metallicity and age results from the SBF and $(B-R)$ measurements are 
valid under the assumption that the studied SBF fields in each galaxy represent the 
dominant stellar populations and that a lower age limit of 8 Gyrs is correct for those 
systems. Within these constraints, they offer a valuable alternative to 
obtain rough metallicities and ages of the unresolved stellar populations 
of early-type dwarf galaxies. 

\placefigure{metalmagplot}

\section{Conclusions}
We have presented the first SBF analysis for 16 dwarf elliptical galaxies
in the Virgo cluster. Our main results and conclusions are:

\begin{enumerate}
\item
Following a similar analysis by Jerjen (2003) for Fornax cluster dEs, we
have shown that the SBF method, by employing a good imager at a 8m-class ground-based
telescope such as FORS@VLT, is an efficient tool to determine accurate distances
to dwarf ellipticals as far out as $25$ Mpc. The semi-empirical calibration 
of the $R$-band fluctuation magnitudes is in most cases straightforward to achieve
by chosing many fields over the face of the galaxy with differing $(B-R)$
colour. The mean uncertainty for an individual galaxy in $(m-M)$ is, to be
conservative, $\approx$ 0.2 mag, or roughly 10\% in distance.

\item
The individual SBF distances of the 16 dEs range from 14.9 to 12.3 Mpc,
proving Virgo cluster membership for all of them. However, the
distance spread is considerable, being roughly twice as large as would 
be expected if the cluster were spherical symmetric. So
even the early-type component of the Virgo cluster is slightly elongated 
along the line-of-sight, confirming similar findings based on the distribution 
of giant Es and intracluster PNe.

\item
The combined giant\,+\,dwarf Virgo early-type SBF distances show a significantly
bimodal distribution. There is a broad primary clump of galaxies at
$D \approx$ 16 Mpc and a narrow secondary clump around 18.5 Mpc. These
features can be associated with the two major Virgo core subclusters centered
on M87 and M86. The latter, smaller subcluster seems to be falling into the 
dominating M87 clump from the backside, separated by ca.~2.5 Mpc. 
A trace of the expected infall pattern is indeed seen in a plot of SBF
distance versus heliocentric radial velocity.

\item
The large depth and bimodality of the distance distribution of Virgo cluster
early types renders a clear-cut definition of the `Virgo cluster mean
distance' difficult. As the M87 subclump, judged from X-ray images, is 
undoubtedly the 
most massive structure of the Virgo complex, it is most natural and
physically meaningful to identify it 
with the Virgo cluster proper. In this case the mean distance of the 
`Virgo cluster' is $\approx$ 16 Mpc, rather than $\approx$ 17 Mpc.

\item
The surface brightness fluctuations of a galaxy also serve as a constraint
on its stellar contents. In particular, the model calibration of the 
observed colour-fluctuation magnitude relation can be used to derive
metallicities for the Virgo dEs that agree with narrow-band
and spectroscopic measurements. The method works if 
the dwarfs are at least as old as 8 Gyr, for which there is independent 
evidence. The SBF metallicies of the present Virgo dEs range from 
[Fe/H] $\approx -1.4$ to $-0.5$.    
     
\item
Given the great potential of the SBF method to determine
the distances of Virgo dEs in an efficient way (low costs in terms
of telescope time) and with 10\% accuracy, it should prove extremely rewarding
for our understanding of the 3D structure of that most nearby cluster
of galaxies, if this work could be extended and a massive observational
campaign be started to get a good fraction of the   
roughly 800 more dwarf ellipticals in the Virgo cluster left.
\end{enumerate}

\acknowledgements
We like to thank the ESO staff of the VLT UT1+UT2 for providing excellent 
quality images in the service observing mode and the anonymous referee for 
carefully reading the manuscript. The authors are grateful to 
the {\em Swiss National Science Foundation} for the financial support of
this research project.

\clearpage


\clearpage 

\clearpage 

\clearpage 

\clearpage 

\clearpage 

\clearpage 

\clearpage 

\clearpage 

\clearpage 

\clearpage 

\begin{deluxetable}{cllccclr}
\tabletypesize{\scriptsize}
\tablecaption{Basic properties of the observed early-type dwarf galaxies in the Virgo cluster. \label{tbl-1}}
\tablewidth{0pt}
\tablehead{
\colhead{VCC} & \colhead{Galaxy}   & \colhead{Morphological}   &
\colhead{R.A.} & \colhead{Decl.}  & \colhead{} & \colhead{$B_T$} &
\colhead{$v_\odot$}     \\
\colhead{Number} & \colhead{Name}   & \colhead{Type}   &
\colhead{(J2000.0)} & \colhead{(J2000.0)}  & \colhead{Subcluster/Cloud} & \colhead{(mag)} &
\colhead{\kms}        \\
\colhead{(1)} & \colhead{(2)}   & \colhead{(3)}   &
\colhead{(4)} & \colhead{(5)}  & \colhead{(6)} & \colhead{(7)} &
\colhead{(8)}     
}
\startdata
0009  &  IC3019      & dE1,N      & 12h09m22.3s &  +13$^\circ59'30''$   & A & 14.04 & 1804 \\ 
0490  &  IC0783      & dS0(3),N   & 12h21m38.8s &  +15$^\circ44'42''$   & A & 13.97 & 1293 \\
0810  &  13o42       & dE0,N      & 12h25m33.6s &  +13$^\circ13'38''$   & A & 16.68  & $-$340 \\
0815  &  13o47       & dE2,N      & 12h25m37.1s &  +13$^\circ08'37''$   & A & 15.95  & $-$700 \\
0846  &  13o48       & dE1,N      & 12h25m50.4s &  +13$^\circ11'52''$   & A & 16.19  & $-$730 \\
0856  &  IC3328      & dE1,N      & 12h25m57.9s &  +10$^\circ03'14''$   & B & 14.19 &  972 \\
0928  &  12o42       & dE6,N      & 12h26m39.6s &  +12$^\circ30'48''$   & A & 16.13 & $-254$\tablenotemark{a}\\
0929  &  NGC4415     & d:E1,N     & 12h26m40.9s &  +08$^\circ26'11''$   & B & 13.69 &  910 \\
0940  &  IC3349      & dE1,N      & 12h26m47.1s &  +12$^\circ27'15''$   & A & 14.81 & 1563\\
1010  &  NGC4431     & dS0(5),N   & 12h27m26.6s &  +12$^\circ17'27''$   & A & 13.86 &  913 \\
1036  &  NGC4436     &dE6/dS0(6),N& 12h27m41.6s &  +12$^\circ18'59''$   & A & 13.89 & 1163 \\
1087  &  IC3381      & dE1,N      & 12h28m15.1s &  +11$^\circ47'23''$   & A & 14.15 &  645\\
1104  &  IC3388      & dE5,N      & 12h28m27.9s &  +12$^\circ49'24''$   & A & 15.49 & 1704 \\
1261  &  NGC4482     & d:E5.N     & 12h30m10.4s &  +10$^\circ46'46''$   & A & 13.59 & 1850 \\
1355  &  IC3442      & dE2,N      & 12h31m20.0s &  +14$^\circ06'54''$   & A & 14.52 & 1332\tablenotemark{b}\\
1422  &  IC3468      & E1,N:      & 12h32m14.2s &  +10$^\circ15'05''$   & A & 13.80 & 1372 \\%
\enddata
\tablenotetext{a}{From Conselice, Gallagher, \& Wyse (2001).}
\tablenotetext{b}{the velocity of 6210\kms quoted for this galaxy in NED is wrong.}
\end{deluxetable}

\clearpage

\begin{deluxetable}{lcccccccccc}
\tabletypesize{\scriptsize}
\tablecaption{Observing log of the imaging data taken in service mode at the FORS+VLT. \label{tbl-2}}
\tablewidth{0pt}
\tablehead{
\colhead{} & \colhead{}   & \colhead{}   & \colhead{Mean} & \colhead{Exptime}  & \colhead{FWHM} & 
 \colhead{}   & \colhead{}   & \colhead{Mean} & \colhead{Exptime}  & \colhead{FWHM}   \\
\colhead{VCC} & \colhead{Filter}   & \colhead{UT Date}   & \colhead{Airmass} & \colhead{(sec)}  & \colhead{(arcsec)} & 
\colhead{Filter}   & \colhead{UT Date}   & \colhead{Airmass} & \colhead{(sec)}  & \colhead{(arcsec)}  \\ 
\colhead{(1)} & \colhead{(2)}   & \colhead{(3)}   & \colhead{(4)} & \colhead{(5)}  & \colhead{(6)} & 
\colhead{(7)} & \colhead{(8)}   & \colhead{(9)}   & \colhead{(10)} & \colhead{(11)}   
}
\startdata
0009 & $B$ &  2000-03-28 & 1.28 & $3 \times 400$ & 0.60 & $R$         &  1999-07-14 & 1.55 & $3 \times 400$   & 0.65\\
0490 & $B$ &  2000-03-28 & 1.34 & $3 \times 400$ & 0.70 & $R$         &  1999-07-10 & 1.41 & $3 \times 400$   & 0.45\\
0810 & $B$ &  2000-05-01 & 1.41 & $3 \times 600$ & 0.90 & $R$         &  1999-07-13 & 1.57 & $3 \times 600$   & 0.60\\
0815 & $B$ &  2000-05-01 & 1.41 & $3 \times 600$ & 0.90 & $R$         &  1999-07-13 & 1.57 & $3 \times 600$   & 0.60\\
0846 & $B$ &  2000-05-01 & 1.41 & $3 \times 600$ & 0.90 & $R$         &  1999-07-13 & 1.57 & $3 \times 600$   & 0.60\\
0856 & $B$ &  2000-03-28 & 1.29 & $3 \times 400$ & 0.85 & $R$         &  1999-07-13 & 1.39 & $3 \times 400$   & 0.55\\
0928 & $B$ &  2000-04-01 & 1.36 & $3 \times 500$ & 0.80 & $R_{\rm s}$ &  2000-04-05 & 1.30 & $3 \times 500$   & 0.60\\ 
0929 & $B$ &  2000-04-05 & 1.19 & $3 \times 400$ & 0.70 & $R$         &  1999-07-14 & 1.47 & $3 \times 400$   & 0.60\\
0940 & $B$ &  2000-04-01 & 1.36 & $3 \times 500$ & 0.80 & $R_{\rm s}$ &  2000-04-05 & 1.30 & $3 \times 500$   & 0.45\\ 
1010 & $B$ &  2000-04-01 & 1.29 & $3 \times 400$ & 0.70 & $R_{\rm s}$ &  2000-04-01 & 1.25 & $3 \times 400$   & 0.65\\ 
1036 & $B$ &  2000-04-01 & 1.29 & $3 \times 400$ & 0.65 & $R_{\rm s}$ &  2000-04-01 & 1.25 & $3 \times 400$   & 0.65\\ 
1087 & $B$ &  2000-04-01 & 1.27 & $3 \times 400$ & 0.65 & $R_{\rm s}$ &  2000-04-01 & 1.24 & $3 \times 400$   & 0.60\\ 
1104 & $B$ &  2000-04-05 & 1.71 & $3 \times 500$ & 0.55 & $R_{\rm s}$ &  2000-04-05 & 1.51 & $3 \times 600$   & 0.45\\ 
1261 & $B$ &  2000-04-01 & 1.43 & $3 \times 400$ & 0.85 & $R_{\rm s}$ &  2000-04-05 & 1.23 & $3 \times 400$   & 0.50\\ 
1355 & $B$ &  2000-03-30 & 1.39 & $3 \times 500$ & 0.90 & $R_{\rm s}$ &  2000-04-05 & 1.40 & $3 \times 500$   & 0.45\\
1422 & $B$ &  2000-04-01 & 1.56 & $3 \times 400$ & 0.75 & $R_{\rm s}$ &  2000-04-05 & 1.21 & $3 \times 400$   & 0.50\\
\enddata
\end{deluxetable}

\clearpage

\begin{deluxetable}{lcccccccc}
\tabletypesize{\scriptsize}
\tablecaption{Quantities derived from the SBF analysis of VCC0009, VCC0490, VCC0810, VCC0815, VCC0846, VCC0856,
VCC0928, and VCC0929.\label{tbl-3}}
\tablewidth{0pt}
\tablehead{
\colhead{} & \colhead{size}   & \colhead{$m_1$}   & \colhead{$\overline{g}$} & \colhead{$s$}  & \colhead{$P_0$} & 
 \colhead{$P_1$}   & \colhead{$S/N$}   & \colhead{$P_{\rm BG}/P_0$} \\
\colhead{Name} & \colhead{(pixels)}   & \colhead{(mag)}   & \colhead{(ADU)} & \colhead{(ADU)} & 
\colhead{(ADU s$^{-1}$ pixel$^{-1}$)}  & \colhead{(ADU s$^{-1}$ pixel$^{-1}$)} & \colhead{} & \colhead{} \\ 
\colhead{(1)} & \colhead{(2)}   & \colhead{(3)}   & \colhead{(4)} & \colhead{(5)}  & \colhead{(6)} & 
\colhead{(7)} & \colhead{(8)}   & \colhead{(9)}  
}
\startdata
VCC0009   F1 & 100 & 27.11 & 1992.9 & 10326.7 & 0.070(0.003) & 0.005 & 10.9 &   0.02 \\ 
\dotfill  F2 & 100 &       & 1393.1 &         & 0.065(0.003) & 0.007 &  7.9 &   0.02 \\ 
\dotfill  F3 & 100 &       &  480.2 &         & 0.082(0.004) & 0.017 &  4.3 &   0.02 \\ 
\dotfill  F4 & 100 &       &  516.7 &         & 0.082(0.006) & 0.017 &  4.5 &   0.02 \\ 
\dotfill  F5 & 100 &       &  681.3 &         & 0.067(0.005) & 0.012 &  4.7 &   0.02 \\ 
\dotfill  F6 & 100 &       & 1412.5 &         & 0.069(0.002) & 0.006 &  8.4 &   0.02 \\ 
\dotfill  F7 & 100 &       &  556.8 &         & 0.090(0.005) & 0.015 &  5.3 &   0.02 \\ 
VCC0490   F1 & 100 & 27.12 &  952.1 & 26184.5 & 0.078(0.003) & 0.017 &  4.1 &   0.02 \\ 
\dotfill  F2 &  80 &       &  404.8 &         & 0.107(0.007) & 0.032 &  3.1 &   0.01 \\ 
\dotfill  F3 & 100 &       &  486.4 &         & 0.098(0.003) & 0.029 &  3.2 &   0.02 \\ 
\dotfill  F4 & 100 &       &  932.6 &         & 0.076(0.003) & 0.017 &  4.1 &   0.02 \\ 
\dotfill  F5 & 100 &       &  766.3 &         & 0.074(0.003) & 0.020 &  3.4 &   0.02 \\ 
\dotfill  F6 &  50 &       & 3568.1 &         & 0.051(0.003) & 0.005 &  7.8 &   0.03 \\ 
\dotfill  F7 & 100 &       &  498.5 &         & 0.082(0.004) & 0.029 &  2.6 &   0.02 \\ 
VCC0810   F1 &  90 & 27.10 &  842.3 & 20052.6 & 0.078(0.005) & 0.013 &  5.1 &   0.02 \\ 
\dotfill  F2 &  60 &       &  870.2 &         & 0.081(0.006) & 0.014 &  5.0 &   0.02 \\ 
\dotfill  F3 &  60 &       & 1243.2 &         & 0.079(0.006) & 0.007 &  9.0 &   0.02 \\ 
\dotfill  F4 &  60 &       & 1153.2 &         & 0.082(0.006) & 0.008 &  8.3 &   0.02 \\ 
VCC0815   F1 &  60 & 27.10 &  942.0 & 20052.6 & 0.077(0.003) & 0.008 &  7.7 &   0.02 \\ 
\dotfill  F2 &  60 &       &  449.8 &         & 0.089(0.006) & 0.016 &  4.9 &   0.02 \\ 
\dotfill  F3 &  60 &       &  653.5 &         & 0.108(0.006) & 0.013 &  7.1 &   0.01 \\ 
\dotfill  F4 &  60 &       & 1277.9 &         & 0.085(0.003) & 0.007 & 10.0 &   0.02 \\ 
\dotfill  F5 &  60 &       &  669.6 &         & 0.091(0.003) & 0.012 &  6.6 &   0.02 \\ 
\dotfill  F6 &  80 &       & 1001.0 &         & 0.076(0.003) & 0.009 &  6.9 &   0.02 \\ 
VCC0846   F1 &  70 & 27.10 & 1236.3 & 20052.6 & 0.084(0.006) & 0.008 &  8.3 &   0.02 \\ 
\dotfill  F2 &  70 &       & 1107.0 &         & 0.075(0.004) & 0.009 &  6.8 &   0.02 \\ 
\dotfill  F3 &  70 &       & 1246.2 &         & 0.059(0.003) & 0.008 &  6.1 &   0.03 \\ 
\dotfill  F4 &  70 &       &  325.2 &         & 0.104(0.007) & 0.025 &  3.8 &   0.01 \\ 
\dotfill  F5 &  80 &       &  981.9 &         & 0.074(0.008) & 0.011 &  5.6 &   0.02 \\ 
VCC0856   F1 & 100 & 27.12 &  176.1 & 10619.7 & 0.120(0.007) & 0.049 &  2.3 &   0.01 \\ 
\dotfill  F2 & 100 &       &  357.5 &         & 0.084(0.004) & 0.028 &  2.8 &   0.02 \\ 
\dotfill  F3 &  70 &       &  583.1 &         & 0.074(0.002) & 0.015 &  4.4 &   0.02 \\ 
VCC0928   F1 &  60 & 27.35 &  503.9 & 10080.5 & 0.073(0.005) & 0.014 &  4.3 &   0.03 \\ 
\dotfill  F2 &  60 &       & 1986.7 &         & 0.062(0.006) & 0.006 &  7.3 &   0.04 \\ 
\dotfill  F3 &  60 &       &  891.6 &         & 0.063(0.004) & 0.009 &  5.3 &   0.04 \\ 
\dotfill  F4 &  60 &       & 2172.5 &         & 0.059(0.004) & 0.004 &  9.1 &   0.04 \\ 
\dotfill  F5 &  60 &       & 2071.7 &         & 0.050(0.004) & 0.006 &  5.7 &   0.05 \\ 
VCC0929   F1 & 120 & 27.12 & 2830.0 & 10408.2 & 0.046(0.002) & 0.005 &  7.2 &   0.03 \\ 
\dotfill  F2 & 120 &       & 3650.1 &         & 0.040(0.002) & 0.004 &  7.0 &   0.04 \\ 
\dotfill  F3 & 120 &       & 1553.6 &         & 0.046(0.003) & 0.008 &  4.7 &   0.03 \\ 
\dotfill  F4 & 120 &       & 1766.1 &         & 0.048(0.003) & 0.007 &  5.2 &   0.03 \\ 
\dotfill  F5 & 120 &       & 1212.6 &         & 0.050(0.003) & 0.009 &  4.4 &   0.03 \\ 
\dotfill  F6 & 120 &       & 1137.7 &         & 0.051(0.002) & 0.010 &  4.4 &   0.03 \\ 
\dotfill  F7 & 120 &       &  696.5 &         & 0.052(0.003) & 0.014 &  3.2 &   0.03 \\ 						   
\enddata
\end{deluxetable}

\clearpage

\begin{deluxetable}{lcccccccc}
\tabletypesize{\scriptsize}
\tablecaption{Quantities derived from the SBF analysis of VCC0940, VCC1010, VCC1036, VCC1087, 
VCC1104, VCC1261, VCC1355, and VCC1422. \label{tbl-4}}
\tablewidth{0pt}
\tablehead{
\colhead{} & \colhead{size}   & \colhead{$m_1$}   & \colhead{$\overline{g}$} & \colhead{$s$}  & \colhead{$P_0$} & 
 \colhead{$P_1$}   & \colhead{$S/N$}   & \colhead{$P_{\rm BG}/P_0$} \\
\colhead{Name} & \colhead{(pixels)}   & \colhead{(mag)}   & \colhead{(ADU)} & \colhead{(ADU)} & 
\colhead{(ADU s$^{-1}$ pixel$^{-1}$)}  & \colhead{(ADU s$^{-1}$ pixel$^{-1}$)} & \colhead{} & \colhead{} \\ 
\colhead{(1)} & \colhead{(2)}   & \colhead{(3)}   & \colhead{(4)} & \colhead{(5)}  & \colhead{(6)} & 
\colhead{(7)} & \colhead{(8)}   & \colhead{(9)}   
}
\startdata
VCC0940   F1 &  90 & 27.35 & 2663.8 & 10080.5 & 0.058(0.002) & 0.003 & 10.2 &   0.04 \\ 
\dotfill  F2 & 100 &       & 1535.7 &         & 0.056(0.002) & 0.005 &  6.9 &   0.04 \\ 
\dotfill  F3 & 110 &       &  661.8 &         & 0.068(0.003) & 0.012 &  4.6 &   0.04 \\ 
\dotfill  F4 &  60 &       &  342.3 &         & 0.116(0.007) & 0.016 &  6.1 &   0.02 \\ 
VCC1010   F1 &  80 & 27.32 & 1303.8 & 8355.3  & 0.054(0.002) & 0.005 &  7.4 &   0.04 \\ 
\dotfill  F2 &  80 &       & 1210.8 &         & 0.052(0.004) & 0.005 &  6.8 &   0.04 \\ 
\dotfill  F3 &  80 &       &  876.3 &         & 0.060(0.004) & 0.008 &  5.9 &   0.04 \\ 
\dotfill  F4 &  80 &       &  660.3 &         & 0.075(0.004) & 0.010 &  6.0 &   0.03 \\ 
\dotfill  F5 &  80 &       &  812.0 &         & 0.064(0.007) & 0.009 &  5.6 &   0.04 \\ 
\dotfill  F6 &  80 &       &  947.3 &         & 0.059(0.004) & 0.007 &  6.2 &   0.04 \\ 
VCC1036   F1 &  60 & 27.32 & 1102.7 & 8355.3  & 0.078(0.004) & 0.006 &  9.6 &   0.03 \\ 
\dotfill  F2 &  60 &       & 1734.3 &         & 0.078(0.003) & 0.005 & 11.0 &   0.03 \\ 
\dotfill  F3 &  60 &       & 1340.8 &         & 0.074(0.003) & 0.005 &  9.7 &   0.03 \\ 
\dotfill  F4 &  60 &       & 2122.0 &         & 0.073(0.004) & 0.004 & 11.7 &   0.03 \\ 
VCC1087   F1 &  70 & 27.31 & 3198.6 & 7903.0  & 0.054(0.002) & 0.002 & 11.1 &   0.04 \\ 
\dotfill  F2 &  60 &       & 3237.6 &         & 0.056(0.002) & 0.003 & 11.4 &   0.04 \\ 
\dotfill  F3 &  60 &       & 2998.2 &         & 0.068(0.002) & 0.003 & 13.0 &   0.03 \\ 
\dotfill  F4 &  60 &       & 2514.5 &         & 0.058(0.002) & 0.003 & 10.4 &   0.04 \\ 
VCC1104   F1 &  90 & 27.33 & 2478.4 & 14213.5 & 0.108(0.004) & 0.006 & 12.0 &   0.02 \\ 
\dotfill  F2 &  90 &       & 1933.2 &         & 0.117(0.002) & 0.007 & 12.1 &   0.02 \\ 
\dotfill  F3 &  90 &       & 1516.8 &         & 0.102(0.003) & 0.009 &  8.9 &   0.02 \\ 
\dotfill  F4 &  90 &       & 1913.6 &         & 0.118(0.004) & 0.009 &  9.9 &   0.02 \\ 
VCC1261   F1 & 110 & 27.37 & 5418.7 & 7873.2  & 0.058(0.002) & 0.002 & 12.6 &   0.04 \\ 
\dotfill  F2 & 100 &       & 4104.7 &         & 0.054(0.001) & 0.002 & 10.9 &   0.05 \\ 
\dotfill  F3 & 110 &       & 1945.4 &         & 0.065(0.003) & 0.005 &  8.8 &   0.04 \\ 
\dotfill  F4 & 110 &       &  983.1 &         & 0.069(0.003) & 0.006 &  7.7 &   0.04 \\ 
VCC1355   F1 &  90 & 27.34 & 2377.9 & 10534.8 & 0.078(0.003) & 0.003 & 13.2 &   0.03 \\ 
\dotfill  F2 &  80 &       & 2640.1 &         & 0.075(0.003) & 0.003 & 12.7 &   0.03 \\ 
\dotfill  F3 &  90 &       & 2360.9 &         & 0.071(0.002) & 0.004 & 11.6 &   0.03 \\ 
\dotfill  F4 &  90 &       & 1834.7 &         & 0.076(0.003) & 0.004 & 11.1 &   0.03 \\ 
VCC1422   F1 & 120 & 27.38 & 3695.1 & 7571.6  & 0.076(0.002) & 0.003 & 13.0 &   0.03 \\ 
\dotfill  F2 & 120 &       & 2209.8 &         & 0.076(0.003) & 0.004 & 10.8 &   0.03 \\ 
\dotfill  F3 & 120 &       & 1018.1 &         & 0.092(0.003) & 0.007 &  9.4 &   0.03 \\ 
\dotfill  F4 & 120 &       &  439.2 &         & 0.109(0.002) & 0.014 &  6.6 &   0.02 \\ 
\dotfill  F5 & 120 &       & 3987.8 &         & 0.066(0.002) & 0.003 & 12.3 &   0.04 \\ 
\dotfill  F6 & 120 &       &  876.2 &         & 0.089(0.004) & 0.008 &  8.5 &   0.03 \\ 
\enddata
\end{deluxetable}

\clearpage

\begin{deluxetable}{cccc}
\tabletypesize{\scriptsize}
\tablecaption{Fluctuation magnitudes and local colours for each SBF field 
in VCC0009, VCC0490, VCC0810, VCC0815, VCC0846, VCC0856, VCC0982, and 
VCC929. All quantities were corrected for Galactic extinction. \label{tbl-5}}
\tablewidth{0pt}
\tablehead{
\colhead{} & \colhead{A$_R$}   & \colhead{$\overline{m}_R^0$}   & \colhead{$(B-R)_0$} \\
\colhead{Name} & \colhead{(mag)}   & \colhead{(mag)} & \colhead{(mag)} \\ 
\colhead{(1)} & \colhead{(2)}   & \colhead{(3)}   & \colhead{(4)}   
}
\startdata
VCC0009  F1 &  $ 0.10 \pm  0.02$ & $29.91 \pm 0.055$ & $ 1.18 \pm  0.03$ \\ 
\dotfill F2 &                    & $29.99 \pm 0.056$ & $ 1.20 \pm  0.04$ \\ 
\dotfill F3 &                    & $29.74 \pm 0.069$ & $ 1.26 \pm  0.08$ \\ 
\dotfill F4 &                    & $29.74 \pm 0.087$ & $ 1.25 \pm  0.07$ \\ 
\dotfill F5 &                    & $29.96 \pm 0.081$ & $ 1.19 \pm  0.06$ \\ 
\dotfill F6 &                    & $29.93 \pm 0.046$ & $ 1.19 \pm  0.04$ \\ 
\dotfill F7 &                    & $29.63 \pm 0.073$ & $ 1.23 \pm  0.07$ \\ 
VCC0490  F1 &  $ 0.06 \pm  0.01$ & $29.86 \pm 0.069$ & $ 1.24 \pm  0.05$ \\ 
\dotfill F2 &                    & $29.50 \pm 0.144$ & $ 1.16 \pm  0.08$ \\ 
\dotfill F3 &                    & $29.60 \pm 0.112$ & $ 1.18 \pm  0.07$ \\ 
\dotfill F4 &                    & $29.88 \pm 0.073$ & $ 1.23 \pm  0.05$ \\ 
\dotfill F5 &                    & $29.91 \pm 0.084$ & $ 1.22 \pm  0.05$ \\ 
\dotfill F6 &                    & $30.33 \pm 0.062$ & $ 1.21 \pm  0.03$ \\ 
\dotfill F7 &                    & $29.80 \pm 0.117$ & $ 1.22 \pm  0.07$ \\ 
VCC0810  F1 &  $ 0.08 \pm  0.01$ & $29.82 \pm 0.087$ & $ 1.23 \pm  0.04$ \\ 
\dotfill F2 &                    & $29.77 \pm 0.091$ & $ 1.22 \pm  0.04$ \\ 
\dotfill F3 &                    & $29.80 \pm 0.092$ & $ 1.25 \pm  0.04$ \\ 
\dotfill F4 &                    & $29.76 \pm 0.084$ & $ 1.26 \pm  0.04$ \\ 
VCC0815  F1 &  $ 0.08 \pm  0.01$ & $29.83 \pm 0.062$ & $ 1.20 \pm  0.03$ \\ 
\dotfill F2 &                    & $29.67 \pm 0.114$ & $ 1.18 \pm  0.04$ \\ 
\dotfill F3 &                    & $29.46 \pm 0.085$ & $ 1.16 \pm  0.03$ \\ 
\dotfill F4 &                    & $29.73 \pm 0.059$ & $ 1.18 \pm  0.03$ \\ 
\dotfill F5 &                    & $29.65 \pm 0.074$ & $ 1.15 \pm  0.03$ \\ 
\dotfill F6 &                    & $29.84 \pm 0.066$ & $ 1.18 \pm  0.03$ \\
VCC0846  F1 &  $ 0.08 \pm  0.01$ & $29.74 \pm 0.087$ & $ 1.22 \pm  0.03$ \\ 
\dotfill F2 &                    & $29.87 \pm 0.068$ & $ 1.22 \pm  0.03$ \\ 
\dotfill F3 &                    & $30.13 \pm 0.062$ & $ 1.26 \pm  0.03$ \\ 
\dotfill F4 &                    & $29.51 \pm 0.140$ & $ 1.17 \pm  0.05$ \\ 
\dotfill F5 &                    & $29.88 \pm 0.115$ & $ 1.26 \pm  0.03$ \\ 
VCC0856  F1 &  $ 0.07 \pm  0.01$ & $29.37 \pm 0.131$ & $ 1.16 \pm  0.08$ \\ 
\dotfill F2 &                    & $29.76 \pm 0.082$ & $ 1.21 \pm  0.05$ \\ 
\dotfill F3 &                    & $29.91 \pm 0.054$ & $ 1.25 \pm  0.04$ \\ 
VCC0928  F1 &  $ 0.09 \pm  0.01$ & $30.14 \pm 0.086$ & $ 1.12 \pm  0.04$ \\ 
\dotfill F2 &                    & $30.33 \pm 0.095$ & $ 1.16 \pm  0.03$ \\ 
\dotfill F3 &                    & $30.30 \pm 0.075$ & $ 1.14 \pm  0.03$ \\ 
\dotfill F4 &                    & $30.38 \pm 0.073$ & $ 1.20 \pm  0.03$ \\ 
\dotfill F5 &                    & $30.56 \pm 0.079$ & $ 1.22 \pm  0.03$ \\ 
VCC0929  F1 &  $ 0.06 \pm  0.01$ & $30.44 \pm 0.056$ & $ 1.40 \pm  0.03$ \\ 
\dotfill F2 &                    & $30.60 \pm 0.059$ & $ 1.41 \pm  0.03$ \\ 
\dotfill F3 &                    & $30.44 \pm 0.064$ & $ 1.39 \pm  0.03$ \\ 
\dotfill F4 &                    & $30.40 \pm 0.066$ & $ 1.40 \pm  0.03$ \\ 
\dotfill F5 &                    & $30.36 \pm 0.061$ & $ 1.39 \pm  0.03$ \\ 
\dotfill F6 &                    & $30.31 \pm 0.057$ & $ 1.40 \pm  0.03$ \\ 
\dotfill F7 &                    & $30.31 \pm 0.070$ & $ 1.38 \pm  0.04$ \\ 
\enddata
\end{deluxetable}

\clearpage

\begin{deluxetable}{cccc}
\tabletypesize{\scriptsize}
\tablecaption{Fluctuation magnitudes and local colours for each SBF field in 
VCC0940, VCC1010, VCC1036, VCC1087, VCC1104, VCC1256, VCC1355 and VCC1422.
All quantities were corrected for Galactic extinction. 
\label{tbl-6}}
\tablewidth{0pt}
\tablehead{
\colhead{} & \colhead{A$_R$}   & \colhead{$\overline{m}_R^0$}   & \colhead{$(B-R)_0$} \\
\colhead{Name} & \colhead{(mag)}   & \colhead{(mag)} & \colhead{(mag)} \\ 
\colhead{(1)} & \colhead{(2)}   & \colhead{(3)}   & \colhead{(4)}   
}
\startdata
VCC0940  F1 &  $ 0.08 \pm  0.01$ & $30.41 \pm 0.051$ & $ 1.32 \pm  0.03$ \\ 
\dotfill F2 &                    & $30.45 \pm 0.047$ & $ 1.31 \pm  0.03$ \\ 
\dotfill F3 &                    & $30.23 \pm 0.058$ & $ 1.30 \pm  0.04$ \\ 
\dotfill F4 &                    & $29.63 \pm 0.086$ & $ 1.16 \pm  0.05$ \\ 
VCC1010  F1 &  $ 0.08 \pm  0.01$ & $30.45 \pm 0.056$ & $ 1.37 \pm  0.03$ \\ 
\dotfill F2 &                    & $30.50 \pm 0.077$ & $ 1.37 \pm  0.03$ \\ 
\dotfill F3 &                    & $30.34 \pm 0.083$ & $ 1.34 \pm  0.04$ \\ 
\dotfill F4 &                    & $30.09 \pm 0.071$ & $ 1.32 \pm  0.04$ \\ 
\dotfill F5 &                    & $30.27 \pm 0.122$ & $ 1.35 \pm  0.04$ \\ 
\dotfill F6 &                    & $30.35 \pm 0.073$ & $ 1.34 \pm  0.03$ \\ 
VCC1036  F1 &  $ 0.07 \pm  0.01$ & $30.05 \pm 0.060$ & $ 1.28 \pm  0.03$ \\ 
\dotfill F2 &                    & $30.05 \pm 0.049$ & $ 1.29 \pm  0.03$ \\ 
\dotfill F3 &                    & $30.10 \pm 0.056$ & $ 1.31 \pm  0.03$ \\ 
\dotfill F4 &                    & $30.12 \pm 0.058$ & $ 1.30 \pm  0.03$ \\ 
VCC1087  F1 &  $ 0.07 \pm  0.01$ & $30.45 \pm 0.042$ & $ 1.39 \pm  0.03$ \\ 
\dotfill F2 &                    & $30.42 \pm 0.044$ & $ 1.38 \pm  0.03$ \\ 
\dotfill F3 &                    & $30.20 \pm 0.042$ & $ 1.36 \pm  0.03$ \\ 
\dotfill F4 &                    & $30.38 \pm 0.050$ & $ 1.36 \pm  0.03$ \\ 
VCC1104  F1 &  $ 0.06 \pm  0.01$ & $29.71 \pm 0.048$ & $ 1.22 \pm  0.03$ \\ 
\dotfill F2 &                    & $29.62 \pm 0.036$ & $ 1.20 \pm  0.03$ \\ 
\dotfill F3 &                    & $29.77 \pm 0.048$ & $ 1.25 \pm  0.03$ \\ 
\dotfill F4 &                    & $29.61 \pm 0.048$ & $ 1.23 \pm  0.03$ \\ 
VCC1261  F1 &  $ 0.08 \pm  0.01$ & $30.42 \pm 0.041$ & $ 1.25 \pm  0.03$ \\ 
\dotfill F2 &                    & $30.51 \pm 0.038$ & $ 1.24 \pm  0.03$ \\ 
\dotfill F3 &                    & $30.29 \pm 0.053$ & $ 1.24 \pm  0.03$ \\ 
\dotfill F4 &                    & $30.23 \pm 0.051$ & $ 1.25 \pm  0.03$ \\ 
VCC1355  F1 &  $ 0.09 \pm  0.01$ & $30.06 \pm 0.054$ & $ 1.32 \pm  0.03$ \\ 
\dotfill F2 &                    & $30.09 \pm 0.057$ & $ 1.34 \pm  0.03$ \\ 
\dotfill F3 &                    & $30.16 \pm 0.041$ & $ 1.34 \pm  0.03$ \\ 
\dotfill F4 &                    & $30.08 \pm 0.050$ & $ 1.34 \pm  0.03$ \\ 
VCC1422  F1 &  $ 0.09 \pm  0.01$ & $30.13 \pm 0.043$ & $ 1.21 \pm  0.03$ \\ 
\dotfill F2 &                    & $30.13 \pm 0.047$ & $ 1.23 \pm  0.03$ \\ 
\dotfill F3 &                    & $29.91 \pm 0.048$ & $ 1.21 \pm  0.03$ \\ 
\dotfill F4 &                    & $29.72 \pm 0.051$ & $ 1.17 \pm  0.05$ \\ 
\dotfill F5 &                    & $30.28 \pm 0.049$ & $ 1.23 \pm  0.03$ \\ 
\dotfill F6 &                    & $29.94 \pm 0.054$ & $ 1.17 \pm  0.04$ \\ 
\enddata
\end{deluxetable}

\clearpage

\begin{deluxetable}{llcc}
\tabletypesize{\normalsize}
\tablecaption{
SBF distances for the dwarf ellipticals (ambiguous cases 
listed twice). \label{distmods}
}
\tablewidth{0pt}
\tablehead{
\colhead{VCC} & \colhead{Galaxy}   & \colhead{$(m-M)_0$}   & \colhead{D} \\
\colhead{Number} & \colhead{Name}   & \colhead{(mag)} & \colhead{(Mpc)} \\ 
\colhead{(1)} & \colhead{(2)}   & \colhead{(3)}   & \colhead{(4)}   
}
\startdata
0009  &  IC3019      &  31.00$\pm$ 0.10  &   15.85$\pm$ 0.80 \\
0490  &  IC0783      &  31.31$\pm$ 0.16  &   18.28$\pm$ 1.46 \\
0810  &  13o42       &  30.76$\pm$ 0.13  &    14.19$\pm$ 0.92 \\
\dotfill & \dotfill     & 31.16$\pm$ 0.22   &   17.06$\pm$ 1.88  \\
0815  &  13o47       &  31.48$\pm$ 0.17   &   19.77$\pm$ 1.68\\
0846  &  13o48       &  31.30$\pm$ 0.19   &   18.20$\pm$ 1.73\\
0856  &  IC3328      &  31.28$\pm$0.25   &   18.03$\pm$ 2.25\\
0928  &  12o42       &  31.42$\pm$ 0.14   &   19.23$\pm$ 1.34\\
0929  &  NGC4415  &  30.86$\pm$ 0.14  &    14.86$\pm$ 1.04\\
0940  &  IC3349      &  31.36$\pm$ 0.20   &   18.71$\pm$ 1.87 \\
1010  &  NGC4431    &  31.05$\pm$ 0.15 &  16.22$\pm$ 1.22  \\
1036  &  NGC4436   &  30.94$\pm$ 0.12  &  15.42$\pm$ 0.93 \\
\dotfill  & \dotfill      &  31.13$\pm$ 0.17 &  16.83$\pm$ 1.43  \\
1087  &  IC3381      &  31.27$\pm$ 0.14   &  17.95$\pm$ 1.26 \\
\dotfill & \dotfill      &  31.39$\pm$ 0.18  &  18.97$\pm$ 1.71   \\
1104  &  IC3388      &  31.15$\pm$ 0.19   &  16.98$\pm$ 1.61\\
1261  &  NGC4482     &  31.34$\pm$ 0.16 &  18.54$\pm$ 1.48\\
\dotfill &  \dotfill     &  31.74$\pm$ 0.21 &  22.28$\pm$ 2.34\\
1355  &  IC3442      &  30.92$\pm$ 0.17   &  15.28$\pm$ 1.30\\
1422  &  IC3468      &  31.64$\pm$ 0.18   &  21.28$\pm$ 1.92\\
\enddata
\end{deluxetable}

\clearpage

\begin{deluxetable}{ccc}
\tabletypesize{\normalsize}
\tablecaption{
$M_V$ magnitudes and median metallicities derived from SBF data.\label{metallist}
}
\tablewidth{0pt}
\tablehead{
\colhead{} & \colhead{$M_V$}   & \colhead{[Fe/H]}  \\
\colhead{Galaxy} & \colhead{(mag)}   & \colhead{(Median)} \\ 
\colhead{(1)} & \colhead{(2)}   & \colhead{(3)}    
}
\startdata
VCC0009         & $-17.68$ & $-1.0$ \\
VCC0490         & $-18.06$ & $-1.0$ \\
VCC0810         & $-15.80/-15.20$ & $-1.1$ \\
VCC0815         & $-16.25$ & $-1.4$ \\
VCC0846         & $-15.83$ & $-1.2$ \\
VCC0856         & $-17.81$ & $-1.2$ \\
VCC0928         & $-16.01$ & $-1.1$ \\
VCC0929         & $-17.62$ & $-0.5$ \\
VCC0940         & $-17.27$ & $-0.7$ \\
VCC1010         & $-17.91$ & $-0.5$ \\
VCC1036         & $-17.77/-17.96$ & $-0.9$ \\
VCC1087         & $-17.84/-17.96$ & $-0.9$ \\
VCC1104         & $-16.38$ & $-1.1$ \\
VCC1261         & $-18.47/-18.87$ & $-1.0$ \\
VCC1355         & $-17.12$ & $-0.6$ \\
VCC1422         & $-18.56$ & $-1.2$ \\
\enddata
\end{deluxetable}

\end{document}